\documentclass[pre,twocolumn,showpacs,amsmath,amssymb,eqsecnum,floatfix]{revtex4}

\usepackage{graphicx}

\usepackage{times}

\begin{document}

\title {Classical height models with topological order}

\author{Christopher L. Henley}
\affiliation{Dept. of Physics, Cornell University,
Ithaca, NY 14853-2501, USA}

\newcommand{\ALT}[1]{}
\newcommand{\HIDE}[1]{}
\newcommand{\MEMO}[1]{}
\newcommand{\THINK}[1]{}
\def\OMIT#1 {{}}   
\def\SAVE#1 {{}}   
\newcommand{\beq}{\begin{equation}}
\newcommand{\eeq}{\end{equation}}
\newcommand{\dagg}{^\dagger}
\newcommand{\Tr}{{\rm Tr}}
\def \la {{\langle}}
\def \ra {{\rangle}}
\def \eqr#1{(\ref{#1})}
\def \rr {{\bf r}}
\def \RR {{\bf R}}
\def \uu {{\bf u}}
\def \si {\sigma}
\def \DU {{\mathcal{D}}}
\def \Group {{\mathcal{G}}}
\def \Auto{{\mathcal{A}}}
\def \Zentrum{{\Group_Z}}
\def \Even {{\mathcal{E}}}
\def \Sel {{\mathcal{S}}}
\def \TC {{\mathcal{T}}}
\def \NC {{\mathcal{N}}}
\newcommand{\Zbb}{{\mathbb{Z}}}
\def \pOK{p_{\rm OK}}   
\newcommand{\smax}{s_{\rm max}}
\newcommand{\nS}{n_\Sel}
\def \pOK{p_{\rm OK}}   
\newcommand{\SP}{S_{\rm Pauling}}
\newcommand{\nG}{{n_{\Group}}}
\newcommand{\nE}{{n_{\Even}}}
\newcommand{\nSel}{{n_{\Sel}}}
\newcommand{\PIest}{{P_1}^{\rm est}}

\begin{abstract}
I discuss a family of statistical-mechanics models 
in which (some classes of) elements of a finite group $\Group$ 
occupy the (directed) 
edges of a lattice; the product around any plaquette is constrained to be
the group identity $e$. 
Such a model may possess topological order, i.e.  its equilibrium ensemble 
has distinct, symmetry-related thermodynamic components
that cannot be distinguished by any {\it local} order parameter.
In particular,  if $\Group$ is a non-abelian group, the topological
order may be non-abelian.
Criteria are given for the viability of particular models, in particular
for Monte Carlo updates. 
\end{abstract}


\maketitle

\section{Introduction}

``Topological order''~\cite{Wen91,Wen04} in a system
means it has an emergent ground state degeneracy 
(in the thermodynamic limit), 
but (in contrast to symmetry-breaking),
no local order parameter operator can distinguish the states.
Topological order has attracted 
great interest over the last 20 years, since (i)  it cannot 
(by definition) be captured by the Landau order-parameter paradigm
and is hence exotic from the viewpoint of traditional solid-state
theory;~\cite{Wen04};
(ii) it is associated with ``fractionalized'' excitations;
(iii) it is proposed to implement qubits by the
ground-state degeneracy, the coherence of which is 
robust against environmental perturbations~\cite{Ki03,Io02};
(iv) the formulation in terms of ground-state degeneracy
makes it attractive for numerical exploration by
exact diagonalization~\cite{Mi05}.

The best-known examples of topological order are
quantum-mechanical: the quantum Hall fluids) and 
lattice models based on $Z_2$, the simplest group,
such as Kitaev's toric code~\cite{Ki03}.
Indeed, Wen~\cite{Wen91,Wen04} once emphasized 
quantum mechanics as a defining property of topological order.
But we can separate these notions: 
topological order (as defined above) is meaningful
in a purely classical model (as developed in this paper)
or in a quantum-mechanical model at $T>0$~\cite{claudios}
(so that e.g. its renormalization-group fixed points represent 
classical behaviors).
Indeed, I would suggest  that the subject of topological order skipped
over more elementary examples, owing to historical accident.
Compare with the history of 
spontaneous symmetry breaking: theorists understood 
classical criticality long before 
quantum criticality~\cite{sachdev}, and
we approach quantum critical properties
in light of their similarities or differences from the classical case.

Analogously, it is hoped that, in the case of topological order,
classical models will (at the least) be a pedagogical aid,
and that behaviors evidenced in classical models may
provide a framework for conjectures about the quantum models.
Disentangling classical notions and inherently quantum  mechanical
ones might lead to clearer (or at least different) thinking.
Also, the framework in this paper naturally draws us to face
hitherto unfamiliar groups -- e.g. the group $A_5$ (see Sec. XX) --
and it might inspire the construction of quantum-mechanical models 
involving these groups.

The explicit notion of ``classical topological order'' was introduced
and highlighted in \cite{claudios}, in particular the points that (i)
it is characterized by ergodicity breaking, (ii) can be implemented by
hard constraints, and (iii) must have a discrete classical dynamics 
-- all of which applies to the models in this paper.  
However, much of Ref.  \cite{claudios} was framed in terms of 
the relation of the classical model to a quantum model, 
e.g. by taking the quantum model to a temperature 
at which quantum coherence is no longer important,~\footnote{
The topological order is spoiled at any $T>0$
in principle, by thermally excited ``vortices'',
since they are deconfined.
However, that is insignificant in
practice if the vortex core energy is made large.
The same would be true in our models if vortices were allowed;
we sidestepped this problem by constraining the configuration 
space to disallow them, in effect setting the core energy
to infinity.}
or by removing some Hamiltonian terms.
In the present work, the model is formulated from the start as
an ensemble of classical statistical mechanics, without concern
for the existence of a quantum counterpart.
That will permit consideration of a richer set of models
(i.e. discrete non-Abelian groups $\Group$), for which we
might not know how to concoct a good quantized version.

I will consider models based on either abelian or non-abelian
discrete groups.  It should be noted that abelianness in
this paper has a different significance than in the quantum
context.  In the latter case, the group in question is the
Berry phase (or its generalization to a unitary matrix)
induced by evolution of the wavefunction from one state
to an equivalent one.   A specific case is the statistics
of quasiparticles whose world lines braid around each other.
Quantum-mechanical non-abelianness  may be realized in 
models based on {\it abelian} groups such as $Z_2$.
Most of the fractional quantum Hall fluids are abelian,
but non-abelian statistics is more suitable for
quantum computation~\cite{Ki03,Fr03}.  
Proposed realizations of {\it nonabelian} 
topological order in this sense are formulated in lattice models as 
sums over loop coverings~\cite{Le05,Fr05}.

\OMIT{[In the abelian case,
one picks up phase factors by exchanging particles
e.g. anyons; in the non-abelian case, 
one produces independent quantum states and the
order of exchanges matters.]}

The primary focus here is not on ideas that point to analytic solutions
or to connections with the existing literature of topological 
order.  Instead, this is meant as a generic blueprint for
numerical studies.  For example, the classification of models
in Sec.~\ref{sec:models} 
(as summarized in the tables) is motivated by the need to
select a good one for simulations, and the quantities defined
in Sec.~\ref{sec:measurements} are all measurable in Monte
Carlo simlations.  However, the actual simulation results
will be left to subsequent papers~\cite{lamberty}.

\subsection{Height models}
\label{sec:height}

I shall realize topological order by generalizing the
concept of ``height models''. 
Their defining 
property~\cite{vB77,Bl82,Ni84,Sa89-Le90,Kon95b,Kon96,Ze97,Ra97,Bu97}
is the existence of a mapping directly from any allowed spin configuration
$\{ \sigma_i \}$ to a configuration of heights $h(\rr)$,
wherein
$h(\rr)-h(\rr')$, for adjacent sites $\rr$ and $\rr'$, 
is a function of the spin variables in the neighborhood
(normally, either two spins on the sites $\rr$ and $\rr'$, 
or else on one spin on site $i$ at the midpoint of the 
$\rr$-$\rr'$ bond: spins and heights commonly live
on different lattices).

The ``spins'' in the model could be any discrete degree of freedom -- 
e.g. dimer coverings. 
For $\{ h(\rr) \}$ to be well defined, it is necessary (and
sufficient) that the sum of height differences is zero round any 
allowed plaquette configuration.  
Thus, a (local) {\it spin-constraint}
is assumed that excludes (at least) the configurations
without well-defined heights, 
yet still allows a nonzero entropy of states 
$S_0$ in the thermodynamic limit.
In the simplest cases, $h(\rr)$ is integer-valued.  

Such models may have ``rough'' phases, in which the coarse-grained
$h(\rr)$ behaves as a Gaussian free field, i.e. the effective
free energy of long-wavelength gradients is
    \beq
    F = \frac{1}{2} \int d^2 \rr |\nabla h|^2 .
    \label{eq:htF}
    \eeq
Via the apparatus of the Coulomb gas formalism ~\cite{nienhuis,nelson-KT}, this 
implies the spin variables have power-law correlations (critical state);
the topological defects may have unbinding transitions like the
Kosterlitz-Thouless transition.  Indeed, in two dimensions
most critical states can be addressed as ``height models''~\cite{nienhuis} 
and this formalism provides an alternative route to computing exact 
critical exponents~\cite{Kon96}, besides conformal field theory.

Note that on a topologically non-trivial space (such as the
torus periodic boundary conditions), there are nontrivial
loops $\ell$, such that the net height difference (or
``winding number'') $w_\ell$ added around such a loop is nonzero.
It is easy to see this is a topological invariant, in that
$w_\ell$ is unchanged if the loop is shifted and deformed (so long 
as it stays topologically equivalent to the original one.)
Thus, if $\ell_1, \ell_2, ...$ are the fundamental loops,
the configuration space divides up into {\it sectors} 
labeled by $(w_{\ell_1}, w_{\ell_2},...)$.
Here, and also for the discrete-group height models 
introduced in the paper, a {\it sector} is each set of 
configurations which can be connected to each other
by a succession of {\it local} spin changes 
(i.e. ``updates'' in the terminology used later).

Point defects may also be admitted and loops around them
may also have a nontrivial $w_\ell$ (in which case they
are topological defects).
Clearly, the winding number $w_\ell$ in a height model is 
analogous to $t\in \TC$ in a topologically ordered model;
we could almost say this is a special case of topological
order in which $\TC$ is $\Zbb$, here meaning the (infinite discrete) 
group of integers under addition.

In place of a Landau order parameter, the (near) degenerate states 
in a topologically ordered system may instead be distinguished
by a global {\it loop} operator, acting around a topologically
nontrivial loop $\ell$. Just as a Landau order parameter
forms a group representation of a broken symmetry, and labels 
the symmetry-broken states, in topological order
the global loop operator ought to form a faithful
representation of the ``topological group $\TC$'' whose
elements label the distinct states.
In our models, the definition of this loop operator is trivial and
transparent: it is just the generalized ``height difference''.

\subsection{Outline of the paper}

In this paper, I first (Sec.~\ref{sec:framework})
generalize the height-model idea
to the case where the height variable belongs to a 
discrete (finite) abelian or non-abelian group, 
thus defining a family of classical models 
which (in many cases) has a topological  order.
The models are defined by a lattice, a group, and the selected
subset of group elements which are permitted values for the 
`spins'' of the model; the spins sit on the bonds.
Non-Abelianness of the group
has interesting consequences: for example, a collection of defects 
no longer has a unique net charge (Sec.~\ref{sec:defects})

In Sec.~\ref{sec:models} and Sec.~\ref{sec:MC}, I survey
the various combinations for the smallest non-Abelian 
groups, using the crude Pauling approximation as a 
figure of merit to identify the most attractive models
for Monte Carlo simulation, using single-site updates.
Furthermore, Sec.~\ref{sec:measurements}
suggests what quantities are  interesting to measure in
such a simulation; however, no simulation results are
reported in this paper.
But some first analytic results are included in
Sec.~\ref{sec:TM}, based on transfer matrices and
hence implicitly one dimensional: the main point
is to shed light on how the size dependence or
defect pair correlation depends on the group 
elements labeling the topological sector or the 
defects.

Finally, the conclusion (Sec.~\ref{sec:discussion})
reflects on which topological behaviors are inherently 
quantum mechanical, and which are not (in that the 
same behavior can be found in classical models).
Furthermore, applications are suggested, either to 
simulating systems with vacancy disorder, or 
to constructing quantum versions of the models 
in this family.

\section{Definitions and topological behaviors}
\label{sec:framework}

This paper is meant to introduce (and compare) a whole family of models.
In this section, I define the general rules for this family,
and then describe the most promising examples.
(In the next section, I shall exhibit consequences for 
Monte Carlo updating of such models.)

\subsection{Model definition}
\label{sec:defns}

Let us take a ``lattice'' (not necessarily Bravais, e.g. honeycomb) 
of sites $\rr$.  The spins sit on the bonds of this lattice, 
and take values in the discrete group $\Group$; they are 
  \beq
      \sigma(\rr, \rr') \in \Group
  \eeq
where $(\rr,\rr')$ labels a bond of nearest neighbors.
The bonds are directed; if we reverse the
direction we view a bond, the spin on it turns into
its inverse:
   \beq
       \sigma(\rr', \rr) \equiv  \sigma(\rr,\rr')^{-1}.
   \label{eq:def-reverse-spin}
    \eeq
Then each configuration of the spins induces a 
configuration of ``heights'' $h(\rr) \in \Group$, defined by
  \beq
  h(\rr)=  \sigma(\rr,\rr') * h(\rr'),
  \label{eq:def-htnonA} 
  \eeq
where ``$*$'' represents the group multiplication.
Of course, $h(\rr)$ is only defined modulo a global 
multiplication by some element $\tau$, $h'(\rr)=h(\rr) \tau$;
to make it well-defined, we could arbitarily require
(say) $h(0)\equiv e$, where $e$ is the group identity.
One could then explicitly construct $h(\rr)$ at the 
neighbors of site $0$, and iteratively at their neighbors,
etc.; the result is independent of which bonds $(\rr,\rr')$
are used for this, if and only if a plaquette constraint
is satisfied [Eq.~\eqr{eq:def-plaq-constraint}, below].

It will be useful throughout to define the {\it line}
or {\it loop} product of $p$ spins:
   \beq
      \gamma(\ell)\equiv \sigma(\rr_1,\rr_p)
      * \sigma(\rr_{p},\rr_{p-1}) * ...
      * \sigma(\rr_2,\rr_1) 
   \label{eq:loop-id}
   \eeq
Here the loop $\ell$ is a string of bonds $(\rr_k,\rr_{k+1})$
connecting end-to-end, for $k=0,...p$; it is a loop when $\rr_p= \rr_0$.

\subsubsection{Plaquette constraint}

Two constraints are imposed on the spin configurations.
The first is the {\it plaquette constraint}:
we require the loop product around 
any elementary plaquette, 
to be the identity, 
    \beq
      \gamma(\ell_{\rm plaq}) = e
      \label{eq:def-plaq-constraint}
    \eeq
This is necessary (and sufficient) for $h(\rr)$ to be
well-defined. 
   
One can define variants of any model by relaxing the 
plaquette constraint to allow a small number of {\it defect} plaquettes, 
around which the loop product is {\it not} the identity.  
Suppose that defects cost an energy $\Delta$ (possibly
depending on the kind of defect): then the basic, defect-free
version of the model can be viewed as a Boltzmann ensemble in 
the limit $T/\Delta \to 0$. On the other hand, if we imagine 
there were a spin-spin interaction that breaks the degeneracy 
of the states satisfying the plaquette constraint, 
then the basic version of the model is the $T/J\to \infty$ limit.

Using condition \eqr{eq:def-plaq-constraint}
and induction (adding one plaquette at a time to
the loop), the loop product must be $\gamma(\ell)=e$ 
for any finite loop $\ell$
that is contractible to a point in small steps.

\subsubsection{Spin constraint}

The second constraint is the {\it spin constraint}: 
choose a ``spin subset'' $\Sel\subset \Group$ such that 
     \beq 
         \sigma(\rr,\rr')\in \Sel
     \eeq
everywhere.  Hence, the  choice of $\Sel$ is a major
part of a model's definition;
in Sec.~\ref{sec:criteria-models}, below, 
I will discuss other desirable features of $\Sel$.
The spin constraint is retained even in versions
of the model with defect plaquettes.

The constraints should respect the group and
lattice symmetries .  Implementing the group
symmetry means requiring
   \beq
    \sigma \in \Sel \Rightarrow  u * \sigma * u ^ {-1} \in \Sel
    \label{eq:def-spin-conjugate}
   \eeq
for any conjugating element $u$.
Thus, $\Sel$ must be one of the group's conjugacy classes 
-- the simplest case -- or a union of such classes.
(Some non-abelian groups, and all abelian ones, 
have ``outer'' automorphisms, symmetries which 
cannot implemented by conjugacy within $\Group$:
we may also wish to implement those symmetries, too).

To implement lattice symmetry, one asks
    \beq
    \sigma \in \Sel \Rightarrow \sigma^{-1}\in \Sel
    \label{eq:def-spin-inverse}
    \eeq
so the model respects inversion (around the bond's midpoint).
~\footnote{In rare cases one drops the condition of including
the group inverse~\eqr{eq:def-spin-inverse}. For example, 
the height representation of the square lattice dimer model
is $\Zbb\{-3,+1\}$sq.}

Let us further require 
    \beq
        e \notin \Sel
    \label{eq:def-spin-identity}
    \eeq
thus a uniform height configuration is disallowed.
Finally, and trivially,
    \beq 
      ~ \Sel ~ \hbox{generates the full group} ~ \Group .
    \label{eq:def-spin-generates}
     \eeq
(If not, I could have redefined $\Group$ as the subgroup 
generated by $\Sel$.)

Without the spin constraint, 
the models would be identical to the lattice
gauge models of Dou\c{c}ot and Ioffe~\cite{doucot-ioffe}.
In such a model, only gauge-invariant (i.e. loop)
quantities can have nonzero expectations;
other correlation functions are zero, even at 
the nearest distance.
In contrast, these group-height models
(like the original height models) have
nontrivial finite-size effects and local correlations:
in particular, there are mediated interactions 
between topological defects.

Furthermore, the spin constraint allows 
the possibility of a long-range ordered phase,
particularly if we assign different Boltzmann weights
to different configurations, and we may find
phase transitions as those parameters aare varied.
Partial orderings are also possible, and 
transitions might occur between different topological orders 
It will be easier to explore the phenomenology of such transitions in the 
classical realm.

\subsection{Topological sectors and topological order}
\label{sec:sectors}

For these models, a ``sector'' means simply the 
configurations that can be accessed by a succession of local updates.
(Here ``local update'' means an operation that turns one
valid configuration to another by changing spins in a small 
neighborhood of some site, as might be deployed for Monte Carlo
simulation.  A ``nonlocal'' rearrangement,
as developed in Sec.~\ref{sec:MC}, 
means the cluster of updated sites can be arbitrarily large, 
and in particular could include a topologically nontrivial
chain of sites that spans the periodic boundary conditions.)
By this definition, ``sectors'' are well defined  in
any finite system larger than the maximum update cluster.
In other models and with other definitions,
passing between sectors is be absolutely forbidden, 
so that sectors (more exactly ``components'', 
like the up and down ordered phases in ferromagnet)
are emergent in the thermodynamic limit.

In  these models, sectors can be labeled by
loop products $\gamma(\ell)$. As noted after 
\eqr{eq:def-plaq-constraint},
products around topologically trivial loops must
give the identity, but others -- e.g.  through the 
periodic boundary conditions of a torus system -- in general
do not.  Such loop products are not changed by local updates, and
therefore must take the same value for all states in a sector.
So we call sectors ``topological'' when they are distinguished 
(and necessarily disconnected) by having different values of
the loop product(s).
The definition of topological order is that
-- in the thermodynamic limit -- different topological sectors
all become equivalent, in that they cannot be distinguished 
by any {\it local} expectations.  Furthermore, just as the
ground state energies in different sectors should become
equal in the case of quantum topological order, the free
energies should become equal in our models.

Let $\ell_1, \ell_2, ...\ell_g$ 
be the basic independent loops, where
$g$ is the genus; then the loop products $\{ \gamma_i\equiv \gamma(\ell_i) \}$
[all taken from the same origin] 
label the possible topological sectors of configuration space.
An interesting question is how many distinct sectors there are,~\cite{oshikawa}
given the system's genus $g$.

\subsubsection{Invariance}

Before counting sectors, we need to explore 
invariance properties of the sector labels,
in case some labels are equivalent to others.
First, there is a sort of gauge freedom:
if we had evaluated these loops starting from 
$\rr$ instead of from 0, then 
    \begin{eqnarray}
       \gamma_\ell  \to \gamma_\ell' &=& \gamma_{0\rr}*\gamma_\ell * \gamma_{0\rr}^{-1},
    \end{eqnarray}
where $\gamma_{0\rr}$ is the line product along any path from the origin to $\rr$;
notice that this same element conjugates {\it all} the distinct loops.
However, just because our labeling fails to distinguish two sectors does
not conclusively show they are the same.

A better criterion for counting sectors as equivalent is that
one can be turned into another by local updates.
Keep the same origin, but perform a single-site update
[see Eq.~eqr{eq:sigma-new-inner}, below]
hitting on the origin vertex, one gets another conjugacy
    \begin{eqnarray}
       \gamma_\ell  \to \gamma_\ell' &=& \tau *\gamma_\ell * \tau^{-1},
   \label{eq:gamma-xy-conjugacy}
    \end{eqnarray}
where $\tau$ is now the updating multiplier.  \SAVE{Yes, I do mean $\tau^{-1}$.}

When the group $\Group$ is {\it abelian}
the sector labels are invariant with respect to how we take the loop 
and unchanged by local updates.
We can have an independent and invariant loop product $\gamma_i$ 
for every topologically independent loop, so the number of sectors 
is $\nG ^{2g}$ where $g$ is the system's genus ($2g=2$ for torus), and
$\nG$ is the number of elements in the group.

\subsubsection{Sector counting in non-abelian case}

\THINK{I need to check what happens when $\Group$ is non-abelian,
but $\TC$ is an abelian subgroup of $\Group$.  Can't a conjugacy due
to a local update change $\gamma(\ell)$ in this case?}

On the other hand, in the non-abelian case, a loop product is
invariant only up to a conjugacy, so we have fewer sectors.
Furthermore, the allowed values of distinct loop products are not independent.
Consider a square lattice model in a rectangular system cell $L_x \times L_y$,
with periodic boundary conditions;
let $(\gamma_x,\gamma_y)$ be the loop products along straight lines 
of bonds running from the origin site $(0,0)$, in the $x$ or $y$ directions respectively.
The loop running from $(0,0)$ to $(L_x,0)$ to $(L_x,L_y)$ to $(0,L_y)$ and back to $(0,0)$
contains no defect, so by inductive use of the plaquette contraint its loop product is $e$.
Yet the four segments of this loop are just $\gamma_x$ and $\gamma_y$, forwards or
backwards, so the loop product is 
   \begin{equation}
       \gamma_y^{-1} * \gamma_x^{-1} * \gamma_y * \gamma_x =e;
   \label{eq:gamma-xy}
   \end{equation}
in other words, {\it $\gamma_x$ and $\gamma_y$ must commute.}

So, in effect, we must define an equivalence relation
$(\gamma_x,\gamma_y) \sim  (\gamma_x',\gamma_y')$ whenever
the pair satisfies \eqr{eq:gamma-xy}, and 
each topological sector is one equivalence class.
In the case of a larger genus, we extend in the obvious way to longer lists;
\SAVE{For the case of periodic BC's in only the $x$ direction, so
$2g\to 1$,  an equivalence class is just a group conjugacy class.}

Some obvious kinds of equivalence classes are:
\par
(i) $(e,e)$
\par
(ii) $(\omega,e)$ or $(e,\omega)$ 
\par
(iii) $(\omega,\omega^k)$ or $(\omega^k,\omega)$ for $k=1,...,m-1$,
where $\omega$ is an element (not the identity) of order $m$.
\SAVE{The exact counting is best done group-by-group.}
\par
(iv) If the group has a nontrivial {\it center} $\Zentrum$, consisting of
elements that commute with all the other elements, then if $(\gamma_x,\gamma_y)$
is a sector then $(z_x \gamma_x, z_y\gamma_y)$ is another sector, where
$z_x, z_y \in \Zentrum$.

Table~\ref{tab:groups} shows the number of classes $\mu_1$ and
the sector count $\mu_2$ for some groups of interest.

\begin{figure}
\resizebox{7.7cm}{!}
{\includegraphics{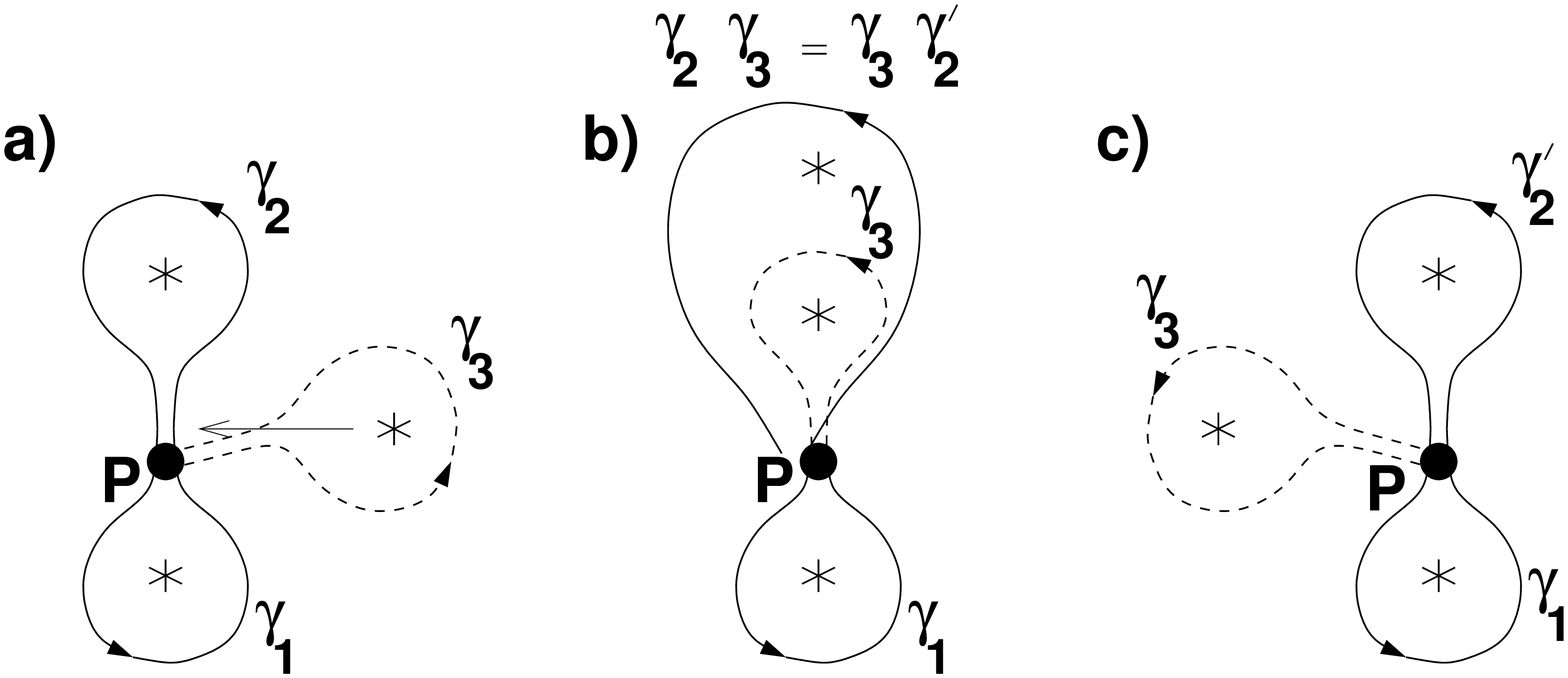}}
\caption{
The composite defect charge of a pair may be changed by
sliding a third defect between the two.
(a). Two defects (stars) have charges $\gamma_1$ and $\gamma_2$,
thus a loop enclosing both contains charge $\gamma_2\gamma_1$.
A third defect with charge $\gamma_3$ is being
moved from the right. 
The loop product around defects 3 and 2 is $\gamma_2 \gamma_3$
These charges are defined using the reference point $P$, and
multiplications are from the right.
(b). After defect 3 is moved between defects
1 and 2, the product around defects 2 and 3 is still $\gamma_2 \gamma_3$.
(c). After defect 3 has passed to the other side,
the product around defect 2 has the new value
$\gamma_2'$; the product around 2 and 3 is $\gamma_3 \gamma_2'$,
which is unchanged from (b) and therefore equal to $\gamma_2\gamma_3$.
Hence $\gamma_2'=\gamma_3 \gamma_2 \gamma_3^{-1}$, and
the combined charge of defects 1 and 2 is now $\gamma_2' \gamma_1=
\gamma_3 \gamma_2 \gamma_3^{-1} \gamma_1$, 
which can  be in a different class than $\gamma_2 \gamma_1$.
}
\label{fig:threedefects}
\end{figure}

\subsection{Defects and non-abelian effects}
\label{sec:defects}

We can allow the possibility (dilutely) of 
a plaquette that  violates the plaquette constraint.
The loop product around it will be called $\beta$. It is analogous
to the Burgers vector of a dislocation, or of the topological defects in the
usual height models based on $\Zbb$).

In the case of a height model (in a rough phase), topological
defects are like vortices in a two-dimensional Coulomb gas\cite{nelson-KT,nienhuis}.
They behave like $U(1)$ electric charges in a two-dimensional universe.
Opposite charges feel an attractive logarithmic potential.
In contrast, in the case of topological order,
the attractive potential decays exponentially and the defects
are deconfined~\cite{moessner-tri-Z2}.  

If the topological order is non-abelian, the
non-commuting property of defect charges has some
interesting consequences.
They are not unique to topological order; this also is a
long known property of defects of traditional ordered states
associated with a non-abelian homotopy group~\cite{mermin}.
One consequence is
that a loop product {\it may} be changed when 
a defect of charge $\beta$ is passed across it, the action being 
a conjugation:
  \beq
      \gamma(\ell)\to \beta\gamma(\ell)\beta^{-1}.
  \eeq
\SAVE{I used the convention of the figure; the defect
being passed is coming near the reference point.}
Thus the topological sector might be changed when a 
defect wanders around the periodic boundary conditions.
Also, the net charge of a defect pair can be changed by passing
another defect between the pair. (See Fig.~\ref{fig:threedefects})

Another consequence of non-abelianness is that 
given two given defects of specified charges, 
there is more than one possible value for their
combined charge. 
In the quantum-mechanical approaches
to defects in topologically ordered systems, this same
property is also the hallmark of non-abelianness.  
In that context, the list of allowed combinations is
known as the ``fusion rules'', and there are matrices 
(generalizations of Clebsch-Gordan coefficients) which tell
how to form the appropriate linear combinations.

A third consequence is pertinent to simulations and the
definition of topological sectors in the presence of
defects.  If one creates a defect pair and moves one defect around 
the boundary conditions, it may recombine with the original defect
into a single defect,  rather than annihilate.
The fact that two defects may not be able to re-annihilate is
very similar to the ``blocking'' idea of Ref.~\cite{oshikawa}
(for quasiparticles in a non-Abelian quantum Hall state).

The single defect state satisfies the generalization of \eqr{eq:gamma-xy}, namely
   \begin{equation}
       \gamma_y^{-1} * \gamma_x^{-1} * \gamma_y * \gamma_x =\beta.
   \label{eq:gamma-defect-xy}
   \end{equation}
This commutation is a ``group commutator''.
Such a single-defect state may conveniently allow numerical
measurements of the creation free energy of a single defect.
Of course, such a state is never possible 
for {\it abelian} defects; in that case,
a system with periodic boundary conditions
must have either no defects, or at least two of them.

\section{Possible models}
\label{sec:models}

In this section, I survey specific models, emphasizing
the criteria which would make some of them particularly
attractive for future investigations.
To summarize Sec.~\ref{sec:defns}:
models in this paper are specified by
(i) the group $\Group$ (values of height)
(ii) the spin subset  $\Sel$ (values of spins)
(iii) the lattice whose bonds the spins sit on.
 
Therefore, the models will be named in the form ``$\Group (m)$latt''.
Here ``$\Group$'' is the groups name, $(m)$ is the
order of the elements in the selected conjugacy class
(usually that is unambiguous), and ``latt'' abbreviates
the lattice (``tri'', ``sq'', or ``hc'' for triangular,
square, and honeycomb). Thus ``$S_3(2,3)$tri'' means
that $\Group$ is the permutations of three objects,
$\Sel$ contains all three pair exchanges,  as well as 
the two cyclic permutations (i.e. every group element 
except for $e$), 
and ``tri'' means the spins sit on the edges of the triangular lattice.
An variant nomenclature is sometimes convenient, in which
the ``$(m)$'' in the label gets replaced by ``$\{\sigma_1, \sigma_2, ...\}$'':
the set $\{\sigma_1, \sigma_2, ...\}$ is simply the listing of the selected
elements.

\subsection{Groups}
\label{sec:groups}

Table~\ref{tab:groups} lists the groups and spin subsets
I shall be interested in.


For future reference, I mention the {\it automorphism group}
$\Auto_\Group$ of a group $\Group$, which is simply its symmetry
group.  Each $a\in \Auto_\Group$ is a permutation of the
group elements preserving its structure, $a(gg')=a(g)a(g')$.
When $\Group$ is non-abelian, there is a
subset of the automorphism group
called the {\it inner automorphisms}, defined as 
the conjugations, $a_\tau(g) \equiv \tau g \tau^{-1}$.
Obviously $a_\tau a_{\tau'} = a_{\tau * \tau'}$, so the
inner automorphism subgroup is isomorphic to $\Group/\Zentrum$, where 
$\Zentrum$ (the center subgroup) consists of the elements that commute with everything.
But many groups have additional {\it outer} automorphisms that
are not conjugations; in particular, all automorphisms of an 
abelian group are outer.

\begin{figure}
\resizebox{8.5cm}{!}
{\includegraphics{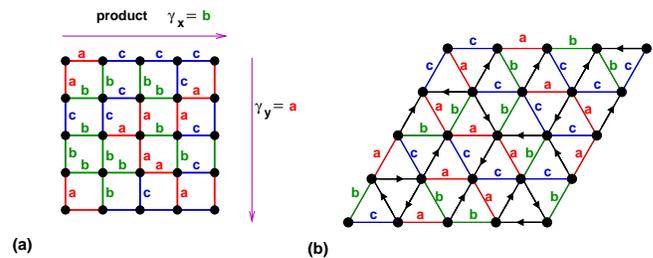}}
\caption{
[Color online.]
(a) Example configuration of the (abelian group) model $Z_2\times Z_2 (2)sq$.
Each square lattice edge is occupied by a group element
$a$, $b$, or $c$; directions are unneeded since each of
these is its own inverse.
The loop products $\gamma_x$ and $\gamma_y$ around the periodic
boundary conditions are also shown, which define the topological
sector. For an abelian group, their values are independent of the
starting points.
(b) Example configuration of the (non-abelian group) model $S_3(2,3)$tri.
Labels $a,b,c$ denote the elements (23),(13),(12), while the
arrow denotes a cyclic exchange (132).}
\label{fig:Z2-example}
\end{figure}

\subsubsection{Abelian groups}

We start by considering discrete abelian groups in this family of models.
The smallest of them, $Z_2$, does not work since $\Sel$ can have
only one element.
The next simplest cases are cyclic groups $Z_q$, i.e.
the integers modulo $q$ under addition, although these
often turn out to be height models (see Sec.~\ref{sec:6-vertex} below).

Beyond that we go to direct products of cyclic groups, indeed
any abelian group can be represented thus.  If $\Sel$ was also
taken to be a direct product, of course the model would
reduce to a superposition of non-interacting models, one
for each factor.   However, there are many attractive examples
in which $\Sel$ is not a direct product, in particular
$Z_2 \times Z_2$ (see Sec.~\ref{sec:3-coloring}).

\subsubsection{Non-abelian groups}

The smallest non-abelian group is $S_3$, the permutations on
three objects (also isomorphic to the dihedral group $D_3$).
Here, $\Sel$ may be taken as the class of all pair
permutations, the model $S_3(2)$, or as all permutations
except the identity, that is $S_3(2,3)$ in our notation.

Each of the next two smallest non-abelian groups has eight elements.
One of these is the 8-element quaternion group $Q$, i.e. the
unit elements $\{\bf \pm 1, \pm i, \pm j, \pm k \}$  from the quaternion ring.
Here $\Sel$ must be the class of the six elements not
equal to $\bf \pm 1$.

The other eight-element non-abelian group is $D_4$,
the symmetry group of the square lattice.

\begin{table}
\caption{Groups and spin subsets.
$\mu$ is the number of conjugacy classes, 
and $\mu_2$ the number of topological sectors on a torus;
$\nG$, $\nSel$, and $\nE$ respectively are the number of 
elements in the group $\Group$, the number in the selected subset
$\Sel$, and the number of even elements.  The effective
bond probability $p_b$ is given by formula~\eqr{eq:pb}.}
\begin{tabular}{@{}lrrrrcl}
\hline 
group+tag        & $\nG$  & $\mu$ & $\mu_2$ & $\nSel$ & $\nE$ &  $p_b$ \\
\hline 
$Z_2 \times Z_2$(2) &   4  &  2 & 16 &   3 &   --  &     1/3  \\
\hline
$S_3$ (2)           &   6  &  3 &  8 &   3 &   3   &     0   \\
$S_3$ (2,3)         &      &    &    &  5  &   --  &     1/5 \\
\hline
$Q$ (2)             &    8 &  3 & 10 & 6 &   --  &     2/7  \\
\hline
$D_4\{m,m'\}$         &    8 &  5 & 20 & 4 &   4   &     4/9?\\
\hline
$A_4$(3)            &   12 &  3 &  8 & 8 &       &     4/11  \\
\hline
$A_5$(2)            &   60 &  4 & 20 & 15 &   --  &     45/59 \\
$A_5$(3)            &      &    &    & 20 &   --  &     40/59  \\
$A_5$(5)            &      &    &    & 12 &   --  &     48/59  \\
\hline 
\end{tabular}
\label{tab:groups}
\end{table}

The ``alternating groups'' $A_4$ and $A_5$ are especially
attractive for our purposes due to their high symmetry
(so we can choose $\Sel$ to be a single class containing
a sizeable fraction of all the group elements).
They consist of the {\it even}
permutations of four and five elements.
Note that $A_4$ and $A_5$ are also the point groups of the
(proper) rotations of a regular tetrahedron and a regular
icosahedron, respectively.
Being subgroups of $SO(3)$, these groups might in some sense
serve as a discretization of it~\cite{ico-hash}, just as clock models are a
discretization of  the XY model.  That would be interesting as
a way to make a connection to topological models
(or gauge theories) defined in terms of Lie (i.e. continuous) groups.

Finally, $A_5$ is the smallest
non-abelian {\it simple} group, meaning it has
no normal subgroups;  as we shall see in a moment
(Sec.~\ref{sec:example-models}), normal subgroups
are an annoyance since they tend to make the behavior
more trivial than would be expected for the group $\Group$.

\ALT{If a group is {\it simple}, meaning it has no normal
subgroups, it has the least risk of reducing to something
more trivial.}

\subsection{Example models}
\label{sec:example-models}

Next I shall survey the simplest examples.
Most of them reduce, in some fashion,
to previously known models; that is an
advantage for computational studies,
since old results can be used as checks.
In several cases, the models in our family
have ``accidental'' topological order,
i.e. beyond the group $\Group$;
In particular, some of them have
height representations.

The group and subgroup 
involved in our spin constraint are finite, and so is
each plaquette;  thus it can happen that the allowed
configurations satisfy stronger constraints than those
they were designed to fulfill.
The first five subheadings below all, in one sense or
another, reduce to known models.

\ALT{
Sometimes the constraints defined by a choice of $\Group$ and
$\Sel$ are actually stronger, in the sense that the same
configurations might represent a different model with a 
larger group.  In particular, it might be a height model.}

\subsubsection{$Z_2\times Z_2$ and the 3-coloring model}
\label{sec:3-coloring}

For a first example, let $\Group \cong Z_2 \times Z_2$,
an abelian group.
Besides the identity, this group has three equivalent elements 
$a$, $b$, $c$; each has order two, and the product of any two 
gives the third.   If we treat these as a class (although they are
not conjugate, since the group is abelian), then we must choose
that class to be the spins, $Z_2 \times Z_2(2)$.  
Since $a=a^{-1}$, etc., we can depict the spins using
three (undirected) ``colors'' of the edges.
On the square lattice this gives perhaps our simplest example
(Figure~\ref{fig:Z2-example}).

What about the triangular lattice case
[model $Z_2\times Z_2(2)tri$]?
The plaquette constraint is simply that each triangle has one 
edge of each color: the ``three-coloring model''. (It is usually 
represented on the edges of the dual [honeycomb], where the
constraint says each vertex has three colors; in either case,
the spins live on kagome lattice vertices, and the configurations
are also the ground states of the 3-state Potts antiferromagnet 
on that lattice.) 
This model is known to have a $\Zbb \times \Zbb$ height 
representation~\cite{huse92,Kon96},
in addition to the finite-group height field $h(\rr)$ defined by 
\eqr{eq:def-htnonA}.

\subsubsection{6-vertex model}
\label{sec:6-vertex}

For another example,
take $\Group$ to be $Z_q$, with $q>4$, and let the
lattice $\{ \rr \}$ be the square lattice.
Choose $\Sel =\{ +1, -1\}$.  (The two elements are not
the same class; they are related 
only by an outer automorphism.)  
Then the sum of spins
around a plaquette can be zero (mod $q$) only if
it is just zero, i.e. there are exactly two $+1$ and two $-1$
in the loop.  If we express these spins on the dual
(also square) lattice,  as an arrow pointing outwards (resp. inwards)
wherever $\sigma=+1$ (resp. $-1$)  as the loop is traversed 
counterclockwise, we see these just are the configurations
of the six-vertex model  -- which also has a integer-valued 
height field.~\footnote{
If we take $\Group$ to be $Z_4$, 
again with $\Sel =\{ +1, -1\}$.  
we get the configurations of the 8-vertex mode.
These correspond 1-to-1 with a set of random spins 
in the dual lattice, so it a trivial $Z_2$ gauge model.}

Since $\Zbb$ or $Z_m$ are abelian groups,  $\{+1,-1\}$ is merely
an {\it outer} class.

\SAVE{
In abelian groups, or in other cases (do they exist?)
that an element $\sigma$ is not conjugate to $\sigma^{-1}$, 
we could define models with a directed spin constraint, 
such that $\sigma$ is allowed only in one direction.  
As mentioned in a footnote, an example is any
dimer covering.
More generally, to accomodate dimer models, 
$\Group=\Zbb$ with
$\Sel = \{ 0, +1\}$ (excluding $-1$) and constrains
the sum around a plaquette to be $+1$ (rather than 
the identity).  The classical dimer model on the triangular 
lattice was pointed out to have $Z_2$ (abelian)
topological order~\cite{moessner-tri-Z2,Mi05}, and indeed it has a 
sort of height representation with $\Group =  Z_2$. 
($Z_p$ with $p$ sufficiently large will do in place of $\Zbb$) or $\Zbb$, chooses
}

\subsubsection {Groups with an even subgroup}

An even subgroup $\Even$ (with $\nE \equiv \nG/2$) 
has $\Group/\Even \cong Z_2$.  That is, any product of an 
even number of elements lies in $\Even$. 
Say that the spin subset $\Sel$ consists of odd elements.
(If it consisted of even elements, we would generate at most
$\Even$.)  
Notice that such a model cannot use the triangular lattice,
since the plaquette rule cannot be satisfied (the plaquette
product must be odd, while $e$ is an even element).

Now if the simulation cell has even dimensions, the possible 
topological products $\gamma(\ell_i)$ must lie in $\Even$.
(Even if the cell has an odd dimension, the possible values of
$\gamma(\ell_i)$ still correspond 1-to-1 with elements of $\Even$.)
Thus, the topological sector labels can only belong to
the subgroup $\Even$.

For example, in permutation groups the
even subgroup consists of even permutations.
In the case of $S_3$, the even permutations are
just $(123)$ and its powers, so $\Even\cong Z_3$.  
Consequently the model
$S_3(2)$sq can have only abelian topological behavior.
In our list, $D_4$ is another group that contains
an even subgroup (the proper rotations).

\subsubsection{Groups with a center}

What if $\Group$ has a non-trivial center $\Zentrum$?
(The center is subgroup of elements that commute with every other element).
For example, if we adopt the group $Q$ of unit axis quaternions, 
(which order 8) then $Q_Z = \{+1,-1\} \cong Z_2$ and
 $Q/Q_Z \cong Z_2 \times Z_2$.
Thus, the model $Q(2)tri$ projects onto configurations of the
the three-coloring model.  
(Just map $\pm {\bf i} \to a, \pm {\bf j} \to b, \pm {\bf k} \to c$.)
The group $D_4$ also has a center (two-fold rotations, i.e. inversions,
commute with everything.)

\THINK{ Well, I believe (need to think through!) that the model is locally
equivalent to the height model defined using the quotient group
$\Group/\Group_Z$, and that is the true topological group  $\TC$.
That quotient has half (or less) as many elements as
$\Group$, so if $\Group$ was one of the smallest non-abelian
groups, the quotient $\Group/\Group_Z$ is probably abelian, 
and in some sense our model has reduced to a more trivial one.
I think the center simply becomes a gauge degree of freedom.}

\subsection{Criteria for models: estimates of entropy}
\label{sec:criteria-models}

To estimate at once the viability of many different models,
I shall use very crude estimates of the entropy and 
(in Sec.~\ref{sec:criteria-update}) updatability.
Say the lattice has coordination $z$ and the dual lattice has 
coordination $z_d$, i.e. the number of sides of each plaquette.
(These numbers are related by $1/z + 1/z_d=1/2$.)
Also, say the group has $\nG$ elements, of which $\nSel$ are in
the selected subset $\Sel$.  These three parameters ---
$\nG$, $\nSel$, and $z$ (or $z_d$) -- contain much of what we
need to characterize the possible models.
See Table~\ref{tab:lattices} for the parameters
related to lattice geometry, and 
Table~\ref{tab:groups} for those related to the groups
and the spin subsets.

\begin{table}
\caption{Lattices 
(asterisk denotes an average over
  two kinds of plaquettes).  
Here ``$\sigma$-phase lattice''
denotes the lattice $(3^2, 4,3,4)$ and ``square-octagon'' lattice
denotes $(4,8^2)$.  The columns give the coordination numbers
$z$ of the lattice and $z_d$ of the dual lattice, 
followed by the lattice's bond and site percolation 
thresholds $p_{cb}$ and $p_{cs}$. 
(Many more significant digits are known~\cite{feng}.)}
\label{tab:lattices}
\begin{tabular}{|l c|clll|}
\hline 
 lattice     & tag  &  $z$ &  $z_d$ & $p_{cb}$& $p_{cs}$\\
\hline 
triangular   & tri  &  6 &  3     &  0.347 &  0.500   \\
$\sigma$-phase  & --   &  5 & 3.333$^*$ & 0.414 & 0.551 \\
square       & sq   &  4 &  4     &  0.500 &  0.593   \\
kagome       & --   &  4 &  4$^*$ &  0.524 &  0.653   \\
honeycomb    & hc   &  3 &  6     &  0.653 &  0.697   \\
square-octagon & -- &  3 &  6$^*$ &  0.677 &  0.730   \\
\hline 
\end{tabular}
\end{table}

I will use a Pauling estimate for the entropy.  There are 
$Nz/2$ edges and hence $\nSel^{Nz/2}$ ways of placing spins
independently chosen from $\Sel$, in a hypothetical ensemble
that does not (yet) enforce the plaquette constraint.
If we knew the fraction of all these states that do
obey the plaquette constraint, we would have the total
count of allowed states and thus the desired entropy.

Pauling's approximation is to pretend
the event of satisfying the plaquette constraint 
is uncorrelated between plaquettes.
So let $f_e$ be the chance that a given plaquette 
has plaquette product equal to $e$.
\SAVE{I called $f_e$ as ``$\pOK$'' formerly.}
Then in this approximation, the probability 
is $f_e^{Nz/z_d}$ that 
the plaquette constraint is satisfied on all $Nz/z_d$ plaquettes
of the whole system.
Thus the Pauling estimate of the ensemble entropy is 
     \beq
         e^{N\SP}  = \Big(\nSel^{z/2} f_e^{z/z_d}\Big)^N .
     \label{eq:SP}
     \eeq
The condition we must satisfy, in order to have an ensemble 
at all, is $\SP>0$, i.e. 
     \beq
     \nSel > 1/f_e^{2/z_d}. 
     \label{eq:SP-ineq},
     \eeq
If $z_d$ is not too small, we may estimate that
the plaquette product is equally likely to be any 
group element, hence {\it very}  crudely 
    \beq
          f_e \approx 1/\nG.
    \label{eq:fe-crude}
    \eeq
Less crudely, one can work out the the actual probabilities 
that the product of $z_d$ random group elements from
the allowed set $\Sel$ will give the identity, 
and these are the $f_e$ values in Table~\ref{tab:models}.
Comparison of those $f_e$ values with 
$\nG$ in Table~\ref{tab:groups} shows that usually,
\eqr{eq:fe-crude} is not bad.  When the product of
two elements of $\Sel$ is particularly likely to fall 
into one class, the true $f_e$ deviates more from
\eqr{eq:fe-crude}, either on the low side
(e.g. the model $A_5(2)$tri) or the high side
(e.g. $A_5(5)$sq).  The extreme case is if the
group contains even and odd elements, and 
$\Sel$ consists of odd elements 
(the $S_3(2)$ or $D_4(m,m')$ examples in Table~\ref{tab:models}).
In that case we should replace \eqr{eq:fe-crude}
by $f_e\approx 1/\nE = 2/nG$ if $z_d$ is even, but $f_e=0$ if
$z_d$ is odd.

There is just one entry in Table~\ref{tab:groups} showing
a {\it negative} Pauling entropy $\SP<0$, namely $A_5(2)$tri.
It is convenient to explain this case in the language of
(proper) rotation group of an icosahedron, which is
isomorphic to $A_5$.  The only way that three twofold elements 
can multiply to give the identity is mutually when the two
fold axes are mutually orthogonal.  Since the triangles
share edges, any valid global configuration must use that
same triad in every triangle; this entails a fivefold ($S_5$) 
global symmetry breaking, since the fifteen twofold axes of
icosahedral symmetry break up into five disjoint orthogonal triads.
Indeed the three used elements form a subgroup isomorphic
to $Z_2\times Z_2$ so we are back at the three-coloring model
for which $\SP>0$.  The Pauling approximation gave zero entropy only 
because it did not take account of the symmetry-breaking and
attempted to mix domains with incompatible symmetry breakings.

My purpose here is {\it not} to obtain quantitative
estimates of the model's entropy, although the Pauling estimate is 
sometimes surprisingly accurate.  Rather, I want to
compare these values between different models as a figure of merit,
to aid us in guessing which models are the most interesting or the most
tractable. To this end, the figures of merit are shown in
Table~\ref{tab:models}.

To satisfy Eq.~\ref{eq:SP-ineq},
the three parameters get pushed in the following 
directions, but there are considerations limiting each of the three.
\begin{itemize}
\item[] 
(1) We want $\nG$ as small as possible; however, there are not 
so many small, discrete, non-abelian groups:  only three 
have $n_\Group \leq 8$,
namely $S_3$ (permutations of three objects), 
$Q$ (quaternion group), or $D_4$ (point group of a square).
\item[] 
(2) We want larger $\nSel$,  meaning the model is less constrained 
(and more tractable). In the limiting case $\nSel=\nG$, the model is
just a pure gauge theory, which is trivial apart from its global
topological  properties.  
On the other hand, a sufficiently large $\nSel$ requires 
including more than one conjugacy class in $\Sel$, so that
the spins can have inequivalent ``flavors''.  That is
esthetically undesirable: a generic model (with unequal statistical
weights) needs more parameters, and it is harder to imagine 
how such a model could be realized physically.  
\item[] 
(3) We want large $z_d$, as in the honeycomb lattice.
However, it is esthetically harder to implement a product 
constraint in a physical model.  (When the product
string is short, there are only a few symmetry-inequivalent
cases for it, and it is easier to concoct a Hamiltonian
term which does not reference the group multiplication, but which
has those cases as its energy minimum.)
\end{itemize}

To satisfy \eqr{eq:SP-ineq} with a large group but $\Sel$ consisting of 
just one conjugacy class, the group must have high symmetry. 
E.g., the alternating group $A_5$ (the proper icosahedral 
rotations) has $\nG=60$ and contains conjugacy classes
with 12, 15, or 20 elements, which using (2) would need
$z_d > 3.30$, $3.02$, or $2.41$ respectively.

\section{Monte Carlo updating}
\label{sec:MC}

For us, one essential criterion of a model is the possibility
of Monte Carlo simulation.  I limit consideration to the
equal-weighted ensemble, in which every allowed configuration
has the same weight.  Then detailed balance is satisfied if
the forwards and backwards rate constants are the same
for any update move.  But what is the minimum sufficient
update move?  For the six-vertex model it sufficed to reverse 
the arrows on the four edges of one plaquette, which changes
the height field on one site, a purely local update.
For the three-coloring model, the minimal update involves
switching two ``colors'' (e.g. $a \leftrightarrow b$) along
a {\it loop}, a nonlocal update move. What happens generically
for our family of models?

\subsection{Cluster update move}
\label{sec:cluster-update}

The update move is simplest described in
terms of the height function
(defined in \eqr{eq:def-htnonA}).
First pick at random a group element $\tau\neq e$
and a starting site $\rr_0$.  Say $\DU$ is the 
domain being touched by the update.
(It will be explained in a moment what determines $\DU$).
Then I prescribe that the 
update premultiplies the heights in this domain by $\tau$, so as to ``shift'' them:
   \beq
          h'(\rr) = \begin{cases}
                     \tau * h(\rr) & \text{for $\rr \in \DU$}; \\
                     h(\rr) &  \text{for $\rr \notin \DU$}.
                    \end{cases}
   \eeq
This induces the following update of the spin configuration:~\footnote{
When both ends $\rr$ and $\rr'$ are in the cluster, 
the mapping $\sigma \to \sigma'$ in \eqr{eq:sigma-new-inner}
is an ``inner automorphism'' of the group, 
as defined in Sec.~\ref{sec:groups}.
The update rule here can be generalized to 
allow outer automorphisms.}
   \beq
          \sigma'(\rr',\rr) = \begin{cases}
                               \tau * \sigma(\rr',\rr)) * \tau^{-1} & 
                                     \text{for $\rr, \rr' \in \DU$}; \\
                               \tau * \sigma(\rr',\rr)) & 
                                \text{for $\rr' \in \DU,  \rr' \notin \DU$}; \\
                                \sigma(\rr',\rr)) * \tau^{-1} & 
                                \text{for $\rr' \notin \DU,  \rr' \in \DU$}; \\
                     \sigma(\rr',\rr) &  \text{for $\rr, \rr' \notin \DU$.}
                    \end{cases}
   \label{eq:sigma-new-inner}
   \eeq
I call this a ``gaugelike'' transformation~\cite{gaugelike}: it has the same
form as a gauge transformation would, but it is valid only when an additional
spin constraint is satisfied too.

If both endpoints of the bond are in $\DU$, then $\sigma'$ is conjugate to
$\sigma$ and must be legal (since we include whole conjugacy classes
in $\Sel$).  

\ALT{
Recall the definition of the height field, \eqr{eq:def-htnonA}:
whenever {\it both} ends of a bond are in $\DU$, we have
$\sigma' = \tau * \sigma * \tau^{-1}$, which
is conjugate to $\sigma$,  and hence 
by    \eqr{eq:def-spin-conjugate} must be legal as $\sigma$ was.}

On the other hand, where the $(\rr,\rr')$ bond crosses the domain 
boundary $\partial\DU$, the spin constraint is nontrivial to satisfy.
Let's place an arrow along the edge from $\rr$  to $\rr'$
if and only if 
   \beq
         \sigma(\rr, \rr') * \tau^{-1}  \notin \Sel .
   \label{eq:arrowdef}
   \eeq
In other words, there is an arrow from $\rr$ to $\rr'$ whenever
including $\rr$ in $\DU$ {\it forces} us to include $\rr'$ as well.
This arrow is not bidirectional.
(it is in the case $\tau^2=e$).
Thus, we might have any of four possibilities 
(no arrows, arrows both way, or arrows one way) along each bond.

\ALT{
On the other hand, 
when only the head $\rr$ is in cluster $\DU$, 
$\sigma'(\rr,\rr') = \tau * \sigma(\rr,\rr')$  
which is not necessarily in $\Sel$.  Hence, I take
$\DU$ to be the {\it smallest} cluster such that every spin on its boundary 
$\partial \DU$ is legal.}

Then the update rule is to construct the arrowed-percolation cluster
consisting of site $\rr_0$, with the rule that site $\rr'$
is included if site $\rr$ is included and there is an arrow
$\rr$ to $\rr'$.
This is the smallest possible
updated domain containing $\rr_0$.
Of course, we do not actually need to construct all the arrows;
instead, we grow the cluster from the initial site, and
construct arrows only from sites already in the cluster.

Notice that (only) in the case $\Group$ is {\it abelian}, 
\eqr{eq:sigma-new-inner} reduces to $\sigma'=\sigma$
throughout the interior of $\DU$. In other words, the
update only changes spins along the boundary $\partial \DU$
and thus is a loop update.  In a non-abelian model,
however,  the update is generally a cluster update.

\ALT{
By this definition, $\DU$ is a percolation cluster for a 
certain correlated, directed percolation problem defined by the instantaneous
spin configuration.  Imagine that, throughout the lattice, 
we draw an arrow on the bond from $\rr$ to $\rr'$ where it would be {\it illegal}
for {\it just} the $\rr$ end to be in $\DU$, i.e., if $\sigma \tau^{-1}$ is 
not in $\Sel$.  (Some bonds get arrows in both directions.)
Then we define $\DU$ such that if the tail site of any arrow is in $\DU$, 
so is the head site; in other words, starting from $\rr_0$, we follow every
outward arrow until we run into a dead end on every site.}

\ALT{Thus we have a cluster update, constructed
as a sort of percolation problem reminiscent of the Wolff updating
algorithm for an Ising model~\cite{wolff}.}

In some models (see next subsection) there is a strong chance to hit a 
system-spanning cluster, including most of the sites, which
tends to be inefficient.  (Updating {\it all} the sites is
equivalent to no update).  To avoid this, a limiting size $\smax$ 
for the update cluster $\DU$ should be set;
if this limit is reached, we cancel the tentative move and start over,
choosing a new random $\rr_0$ and $\tau$.

\subsection{Numerical criteria for cluster updates}
\label{sec:criteria-update}

Notice that in growing a cluster from $\rr_0$,
we never cared about the reverse arrows. 
Therefore, we obtain the same clusters as we would
in an ordinary (not arrowed) percolation problem,
if the occupied bond probability $p_{b}$ is identified with the
probability of an arrow in a pre-selected direction;
that probability is simply 
    \beq
      p_b \approx 1 - \frac{\nSel-1}{\nG-1},
    \label{eq:pb}
    \eeq
if we chose the candidate updating factor $\tau$ at random.
These probabilities are shown in Table~\ref{tab:models}. 
In a case that {\it any} group element 
$\tau$ works as a multiplier on {\it any} bond,
I would write $p_b=0$ in Table~\ref{tab:models}, rather than
use \eqr{eq:pb}.
In such a case, our model is locally trivial:
exactly $\nG^N$ configurations may be accessed, 
simply by applying one arbitrary group element at
every site. In other words, there is a (locally) 1-to-1 mapping to
the trivial model in which every site has an 
independent degree of freedom.
That model is just the gauge model, which was studied 
previously in Ref.~\cite{doucot-ioffe}.

In the spirit of the Pauling approximation, let us now pretend 
that arrows on different bonds are {\it uncorrelated}:
within that assumption, we must obtain the {\it same} cluster distribution 
as in the (thoroughly studied) problem of uncorrelated percolation on
these lattices.
It follows that the updating behavior tends to depend on the relation of 
$p_b$ to the critical percolation fraction $p_{cb}$.  
On the one hand, if $p_b>p_{cb}$, then
the cluster grows without limit, including a nonzero fraction
of the whole system; in that case, the update move certainly
is not efficient.  On the other hand, if $p_b/p_{cb}$ is too
small, we never get a cluster at all, or else a single-site
update (next subsection) would suffice.  The ``interesting'' case
when a cluster update is necessary and helpful, would be for
$p_b/p_{cb}$ close to or slightly less than unity.

\SAVE{Such large-cluster updates are efficient for the updating
Markov process, in that the new configuration is far from 
the old one.  (This is the basis for the Wolff algorithm
and other cluster-update methods for accelerating Monte Carlo.)}

\SAVE{
Notice that even if some selections of $\rr_0$ and
$\tau$ lead to a spanning cluster of this sort,
other choices will lead to finite clusters and allow
the possibility of updating.
However, if $\smax$ is too small, 
the update might be insufficient to access the full 
ensemble; in this situation, we would be rearranging
disconnected parts of the configuration and not really
generating a new one.  For example, in the 3-coloring
model, the smallest possible update ($\smax=1$) is changing
$ababab \to bababa$ along a hexagon; this is not believed
to access all sites.}

\begin{table}
\caption{
Entropy and updatability parameter estimates for selected models.
Formulas from eqs. \eqr{eq:SP}, \eqr{eq:pb}, and \eqr{eq:P1est}.
Note $a$: in these cases, any even element $\tau$ can always update, but no odd $\tau$
can ever update.
}
\begin{tabular}{|l| rrll|}
\hline 
Model name   &          $f_e$ & $\exp(\SP)$  &$p_b/p_{cb}$ & $\PIest$ \\
\hline  
$Z_2 \times Z_2$(2)tri  & 2/9 & $4/3$ &   ...  &   0.241 \\
$Z_2 \times Z_2$(2)sq  &  1/3 & $7/3$ &   0.667  & 0.681 \\
\hline
$S_3$(2)sq             &  1/3 & $3$   &   0.0   & 0.5$^a$ \\
$S_3$(2)hc             &  1/3 & $3$   &   0.0   & 0.5$^a$ \\
$S_3$(2,3)tri          & 4/25 & $16/5$ &   0.576 & 0.870 \\
$S_3$(2,3)sq         & 21/125 & $21/5$ &   0.400 & 0.963 \\
\hline
$Q$(2)sq             &  7/54  & $14/3$ &   0.571   & 0.930 \\
\hline
$D_4\{m,m'\}$sq          & 1/4 & $4$  &   0.0 & 0.5$^a$ \\
$D_4\{m,m'\}$hc          & 1/2  & $4\sqrt2$  &   0.0 & 0.5$^a$ \\
\hline
$A_4$(3)tri           & 1/16 & $2$  &  0.727  & 1.000   \\
$A_4$(3)sq           & 3/32 & $6$  &   0.613  & 1.000   \\
\hline
$A_5$(2)tri           & 2/225 & $4/15$  &  2.20  & 0.668   \\
$A_5$(3)tri           & 7/400 & $49/20$ &  1.95   & 0.894   \\
$A_5$(5)tri           & 5/144 & $25/12$ &  2.34   & 0.706  \\
$A_5$(2)sq        & 71/3375   & $19/5$  &  1.53   & 0.285  \\
$A_5$(3)sq        & 147/8000  & $147/20$&  1.36  & 0.544 \\
$A_5$(5)sq        & 53/1728   & $53/12$ &  1.63  & 0.360   \\
\hline   
\end{tabular}
\label{tab:models}
\end{table}


\subsection{Single-site updates}
\label{sec:singlesite}

A {\it single-site update} is the case that the updated
cluster $\DU$ is just one site, thus only the
$z$ spins around it are updated.
\SAVE{This can be thought of as the smallest loop update move.}
When $p_b$ is much less than $p_{cb}$, most clusters
are small, and the probability $P_1$ of a single-site update 
is appreciable.
If $P_1$ is large enough, it {\it might} be ergodic to use 
{\it only}  single-site updates (i.e. to pick $\smax=1$),  
in which case we can omit the cluster-growing algorithm.   
I shall concentrate on these cases, which are the easiest
to simulate (and also the likeliest to extend to quantum models).

To estimate $P_1$, I pretend the $z$ bonds around a site are
independently occupied by randomly chosen elements of $\Sel$.
Then
   \beq
    \PIest = 1- \prod_\alpha (1-q_\alpha ^z)^{n_\alpha}
    \label{eq:P1est}
   \eeq
where $\alpha$ indexes the group class (with $n_\alpha$ elements)
that $\tau$ might be in (excluding the identity), and 
I defined  $q_\alpha$ to be the fraction of times 
$\tau *\sigma \in \Sel$ given that $\tau$ 
falls in class $\alpha$.

To digest the implications of \eqr{eq:P1est}, let's
make an even cruder version of the estimate, 
replacing $q_\alpha$ by it's average over all $\tau's$,
namely $q_\alpha \to 1-p_b$:
I get
$1- [1-(1-p_b)^z]^{\nG-1}$ which is a lower bound on 
$\PIest$ as given by \eqr{eq:P1est}.
Evidently, to have a high single-site success rate, we 
want (i) $\nG$ as large as possible,
(ii) $z$ as small as possible,
and (iii) $p_b$ as small as possible; in light of
\eqr{eq:pb}, the third criterion amounts to
wanting $\nSel/\nG$ as large as possible.
Those are the same three considerations given
in Sec.~\ref{sec:criteria-models}
as favoring a large entropy.

\ALT{
We arrive at the
rule-of-thumb that (i) the group should be as small
as possible (ii) $\Sel$ should consist of its largest class.
(In the case of $D_4(2,2)$, 
$\Sel$ combines two classes which are equivalent but only
through an outer automorphism.)
}

\SAVE{I find it somewhat mysterious why comparing $\PIest$
to $p_s$ doesn't always tell us whether we have a single
site update.  Consider two height models which are also
special cases of this set-up but using abelian groups.  
For the 6-vertex model on the square lattice, $p_b=1/2$ 
and $\PIest \approx 1/8$; 
for the 3-coloring model on the triangular lattice, 
$p_b= 3/4$ and $\PIest \approx [1-(3/4)^6]^4 \approx 1/11$.}

I include these estimates in Table~\ref{tab:models}, 
particularly focusing on the models using group $A_5$.
We see from Table~\ref{tab:models} that $\PIest$ is large
enough in many cases that we can rely on single-site updates.
However, whenever $\PIest$ gets close to 1, our 
model is ``too easy'' in some sense -- it is 
practically a gauge model, with only mild constraints
eliminating some of the configurations.

The entry $\PIest=1$ for $A_4(3)$ is delusory.  This comes
because $q_\alpha=1$ for a certain class of update multipliers,
namely the order-2 class (double pairwise exchanges).  
If we limited ourselves to this class, indeed 
every update would be successful, but (it can be checked)
the move would not be ergodic (does not access the whole ensemble).
A similar situation applies in the cases of $S_3(2)$ or
$D_4\{m,m'\}$: any $\tau$ from the even subgroup is always
accepted, while an odd $\tau$ is never accepted; but 
single-site updates based on the even subgroup do not
access the whole ensemble.

To implement an actual simulation, one would not want to choose
$\tau$ at random, but biased towards the group classes with a
larger $q_\alpha$ (the success fraction looking at just
an isolated bond).  In particular, one would omit group
classes with $q_\alpha=0$; if the group contains even/odd
elements and $\Sel$ includes only one parity of element, 
then $q_\alpha=0$ for every class of odd elements.  
The values of $p_b$ and $\PIest$ in Table ~\ref{tab:models}
for $S_3(2)$ and $D_4(m,m')$ were computed assuming $\tau\in \Even$.

Another way to implement a single-site update is, after choosing
a random vertex $\rr$, to examine the local environment of its
$z$ bonds, find the entire list of $\tau$'s which can update it,
and choose randomly from this list.  Typically, configuration
dependent choices like this are avoided in Monte Carlo algorithms
because they tend to violate detailed balance.  In the present
case, however, it can be checked that the number of possible 
$\tau's$ is always the same in the old and new configuration,
i.e. the rate is the same for the forward and backward step,
which is sufficient to ensure detailed balance (and an equal
ensemble weight for every configuration).

\subsection
{Criteria for initial conditions}

In height models, certain special states 
(e.g. the ``columnar'' arrangement of dimers on the square
lattice) were ``ideal'' in having the maximum number of
possible update moves.  (For a model requiring loop updates, we
might replace that criterion by ``having the shortest typical
loops.'') Certain other states (e.g. the ``herringbone''
packing of dimers) were inert, in that no finite updates are
possible (in the thermodynamic limit).  These states, in a
height model, correspond respectively to 
a zero coarse-grained gradient of the height variable, 
or the maximum gradient.

In a non-abelian height model, the coarse-grained height gradient
is undefined, but one can still construct ``ideal'' and ``anti
ideal'' states.   It is recommended that simulation runs be  started
in both kinds of state, being in some sense opposite extremes of
the configuration space.  A diagnostic for equilibration is then 
whether the expectations from the two starts converge to the same values.

More exactly, rather than a single domain of anti-ideal state, one should
divide the system into two domains. Then, updates are initially possible 
along the domains' border, using loops which extend across the system. 
Gradually, a larger fraction of the system's area become updatable, 
and the loops get smaller.  On the other hand, starting from an ideal
state, the loops are initially small and get larger.  Thus, tracking the
loop distribution is an obvious diagnostic to test for convergence to
the same equilibrium state.

\section{Possible measurements in simulations}
\label{sec:measurements}

In this section, I sketch how one might 
confirm the topological order numerically,
or measure other interesting quantities,
given a working Monte Carlo simulation.

\subsection{Correlation functions}

Correlation functions are an obvious starting point.
Of course, a topological order state has exponentially
decaying correlations, so this serves primarily as a negative test:
we check that the system is not a height model in disguise
(see Sec.~\ref{sec:example-models}), which would have
power-law correlations, and that it doesn't have long-range order
(which can emerge even in equal-weighted entropic ensembles, 
or because the defining constraints are too restrictive).
Correlations are also of interest near a critical point where long-range
or quasi-long-range order emerges.

In models with vector spins ${\bf s}_i$, one was accustomed 
to evaluating the expectation of ${\bf s}_i \cdot {\bf s}_j$, 
or occasionally its second moment.
It may not be immediately obvious what to measure now.  
One can, of course, simply tabulate frequencies of different
combinations, e.g. (for the ``height difference'') how often
$\gamma_{0\to \rr}$ belongs to each conjugacy class.
It is preferable, though, to reduce the measurements to a 
single (meaningful) number, and the appropriate generalization
of the dot product is the trace of the matrices
in the right group representation.

Thus we are led to use a character function $\chi(x)$, where $x$ is
any group element;  this is always the same within each conjugacy
class of the group.  I divide the actual character by the dimension
of the representation, so that $\chi(e)\equiv 1$ for any 
representation, and $|\chi(x)|\leq 1$ for any element.
Presumably, the best choice of representation is the one that
has the largest positive $\chi(\sigma)$ for spins (for $\sigma\in \Sel$).
This corresponds conceptually to using a distance metric, 
within the group $\Group$, counting many multiplications by 
some element of $\Sel$ are needed to take you from element 
to the other one.

\subsubsection{Height difference correlation}

In the old ``height models'' (sketched in Sec~\ref{sec:height}),
a natural measure of fluctuations was $\langle |h)0)-h(\rr)|^2\ra$.
The natural generalization of this for the present models with
finite (possibly non-abelian) groups is
    \beq
           C_h(\rr) \equiv  \langle \chi (\gamma(\ell_{0\to \rr}))  \rangle .
    \eeq
Of course, the product $\gamma(\ell_{0\to \rr})$ is independent
of which path is taken from $0$ to $\rr$ -- provided the path 
does not wrap around the periodic boundary conditions.

As just noted, choosing $\chi(.)$ so that $\chi(\sigma)$ is as close
to one as possible, provides that $C_h(\rr)$ does
express how fast the group element wanders from
the identity under repeated compositions; that is the choice
likeliest to give a monotonic decay with distance.  
If $\gamma(\ell_{0\to \rr}))$ is equally likely 
to be any group element
-- which  one expects large $\rr$ -- 
then it follows that $C(\rr)= 0$.

\subsubsection{ Spin correlations}

Similarly, we can compute 
    \beq
          G_{ij} \equiv \langle \chi( \sigma_i * \sigma_j^{-1})\rangle  .
    \eeq

\subsection{Defects}
\label{sec:measure-defects}

It is easy to augment the simulation to allow a defect plaquette
where the plaquette constraint is violated.  
The same (single-site) update rules will work correctly next to the defect, 
but they cannot change its position.  To make a defect mobile,
one can add additional update rules specific to the defect, by (say)
arbitrarily choosing one bond of the plaquette and changing it to
make the plaquette's loop product be $e$ (which, of course, the
loop product {\it not} be $e$ for the plaquette on the other
side of that bond, unless that was also a defect plaquette and
this is the annihilation event.)  The simulation would normally
be run with a constraint or bound on the number of defects.

The idea is to create a pair of defects, by hand, and then 
evaluate expectations depending on them.
The first thing to measure is the distribution 
$P_d(\RR)$ of defect separations $\RR$.
In the case of topological order, we expect deconfinement, 
meaning  $P_d(\RR) \to {\rm const}$ for $\RR > \xi$, where
$\xi$ is a (not very large) correlation length. 
One can define an effective (entropic) potential 
$V(\RR)$ by
    \beq
           P_d(\RR) \propto \exp(V(\RR));
    \eeq
physically, $V(\RR)$ is the difference in entropy due to placing the
defects near to each other.  In the case of a height model,
$V(\RR) \propto \ln |\RR|$, and $d_P(\RR)$ decays to zero as a power law.~\footnote{
Differing varieties of confinement behavior are possible,
depending whether that power is fast enough for $\int d^2\RR P(\RR)$ to
converge at large $\RR$.
\SAVE{Most height models have a dimensionless stiffness smaller than the 
critical Kosterlitz-Thouless value, so they will be disordered by
defects if even a small defect fugacity is allowed.}
}
\ALT{We can measure $P_d(\rr)$, the probability of a separation $\rr$.
This is expected to approach a constant, indicating deconfinement
of the defects (in a standard height model it decays as a 
power law).}

In fact, since there are various flavors of defect labeled
by different group elements $b$, one really needs to write the effective
potential as 
    \beq
                 E = U(b)+U(b')+ V_{b,b',c}(\RR)
    \label{eq:defect-int-bbc}
    \eeq
where $b$ and $b'$ are the respective
defect charges, and $c$ is the net charge of the combined defect.
Here, $U(b)$ and $U(b')$ are ``core energies'' of these respective defects;
these, and {\it usually} the inter-defect potential,
are functions only of the conjugacy classes of $b$, $b'$, and/or $c$.
\SAVE{We can imagine, I suppose, groups in which there are inequivalent
ways to combine two elements from the classes of $b$ and $b'$ to make 
an element of class $c$ -- inequivalent in that we can't turn one triple
into the other by conjugacy.}
Implicit in the form \eqr{eq:defect-int-bbc} is that the exponential
confinement length probably depends on all of $b$, $b'$, and $c$. 

Measuring how the effective potential depends on class is more
physical, since (i) it decides whether a defect is stable against
decays into other defects (ii) measurements of defect behavior  
(in simulations or in real systems, were any to be discovered)
might be used to discover the universality class of the topological
order, if that were not known.  
I conjecture that the dependence on $b$, $b'$, and $c$ is also
described by a character function; it would be interesting to
see if that can be explored analytically in some model.

Incidentally, since (with the appopriate boundary conditions)
we can have a {\it single} defect in our system cell, that gives
additional opportunities to evaluate e.g. the core energy
$U(b)$ without the complication of a second defect.

Finally, if the single-site updates of Sec.~\ref{sec:singlesite}
are not feasible, defects provide a less elaborate alternative 
update move, in place of the cluster update of Sec.~\ref{sec:cluster-update}.
Namely, we create a pair of defects and allow them to random-walk
until they annihilate.  (However, if their paths differ by a loop
around the periodic boundary conditions, they may be unable to
annihilate.) 
Many Monte Carlo schemes~\cite{oxborrow,krauth}
are based on a similar process.

\THINK{
Is it possible this only works when a loop update would suffice? (The loop
around the system being the track taken by the two defects to re-annihilate.)
I think it does work generally, but haven't proven it to myself.}

\subsection{Topological sectors}
\label{sec:measure-sectors}

The tests of topological order outlined up to here
have been negative; none of them catpures the {\it positive} 
property of topological order, which is the degeneracy
of topological sectors.  This can be measured in a classical simulation,
if we use a (necessarily nonlocal) update which can change sectors,
while satisfying the detailed balance condition.
Either the cluster update of Sec.~\ref{sec:cluster-update}
or the defect-pair update just outlined in Sec.~\ref{sec:measure-defects}
will suffice.

From the relative fraction of time spent in different topological 
sectors, we can infer a free energy $F_L(\gamma_x,\gamma_y)$, 
where $(\gamma_x, \gamma_y)$ are the loop products characterizing
the sector, and $L$ refers to the system size.   This is a finite
size effect, since (by definition of topological order) the 
difference between sectors vanishes in the thermodynamic limit;
$F_L$ is expected to decay exponentially as a function of $L$. 

In a similar fashion, if we allow transitions between states
with and without a defect as part of the dynamics, we can 
evaluate the defect core energy $U(b)$.
Of course, $F_L(\gamma_x,\gamma_y)$ is very analogous to $U(b)$, 
since $b$ is a loop product encircling the puncture where a defect sits, 
just as $\gamma_x$ is from the loop product encircling the system.
~\footnote{
Indeed, a conformal mapping by the complex logarithm function 
converts the puncture geometry into a strip with periodic boudary
conditions across it.}


\section{Transfer matrix and analytic approaches}
\label{sec:TM}

\newcommand{\Ttilde}{{\widetilde{T}}}

In a quantum mechanical models with topological order, 
the energy differences between different topological sectors 
decays with system size as $\exp$(const~$L$), and the
correlations of a defect pair decay as 
$\exp(-R/\xi)$. Up to now, I have assumed without 
justification that this would carry over to the 
present classical models.

\SAVE{In fact, the origin of this dependence in the quantum 
case seems to be tied up with how the action of Wilson
loops (or Polyakov loops) scales in $(d+1)$-dimensional
path-integral for the phase in question.}

This section finally examines the basis of exponential behavior.
I turn here to an analytic treatment
using the (practically) one-dimensional framework
of transfer matrices.  First of all, this sheds
some light on why the finite-size dependences, as well
as the defect-defect interaction, are exponentially
decaying with distance.  More specifically, they
clarify the pattern of how sector-weight splittings or defect-pair distributions
relate to the group's representations and symmetries.

\subsection{One-dimensional model}
\label{sec:1D-TM-sector}

Imagine the most trivial system which can have
topological sectors: the one-dimensional version of
the discrete-group height models.  There can be no plaquette
constraint. Our ensemble simply consists of chains
of length $L$ -- with periodic boundary conditions --
having a group element $\sigma_i$ placed on each link,
the only constraint being that $\sigma \in \Sel$.
All $(\nS)^L$ sequences are equally likely.

If we let 
   \beq
           \gamma(x) \equiv \sigma_x * \sigma_{x-1}* ...* \sigma_1.
   \eeq
then the topological sectors are labeled by $\gamma(L)$.
Define the ($\nG\times \nG$ dimensional) {\it transfer matrix} $T$ 
in the standard fashion: let $T_{\gamma',\gamma}$ be the number of ways to get
$\gamma(x+1)=\gamma'$ from $\gamma(x)=\gamma$.
Then $(T^L)_{\gamma,e}$ is the partition function
(the total number of states) for the sector with $\gamma(L)=\gamma$.
Note that $T$ commutes with permutations that implement the
symmetry operations (automorphisms) of the group $\Group$; hence,
the eigenvalues/eigenvectors of $T$ are classified by the
representations of the automorphism group (mentioned in
Sec.~\ref{sec:groups}).
The transfer matrix has eigenvalues $\{ \Lambda_k \}$ and corresponding
eigenvectors $\{ v_{k,m} \}$; the index $m$ labels each family 
of symmetry-related eigenvectors belonging to the same (degenerate)
eigenvalue $\Lambda_k$.

The restricted partition function for topological sector $\gamma$ is
   \beq
       \sum _{k,m} [v_{k,m}]_\gamma [v_{k,m}]_e \Lambda_k^L .
   \label{eq:Z-rest-gamma}
   \eeq
Hence, in any sector the overall (entropic) free energy per unit length is
$\ln \Lambda_0$, where $\Lambda_0$ is the largest eigenvalue, 
and $L$-dependent corrections depend 
on some larger eigenvalue $\Lambda_s$.  
For this trivial one-dimensional model, $v_0 = (1,1,..., 1,1)/\sqrt{\nG}$.  
More generally $v_0$ must be totally symmetric under all automorphisms
of $\Group$, i.e. it belongs to the trivial representation.
Indeed, $v_s$ must {\it also} belong to the trivial representation, 
since $[v_{k,m}]_e$ in \eqr{eq:Z-rest-gamma} is independent of $m$, 
but $\sum_m [v_{k,m}]_\gamma =0$ for any other representation.
We let $\Lambda_s$ be next largest (necessarily nondegenerate) 
eigenvalue of the fully symmetric representation, after $\Lambda_0$.

Hence, 
   \beq
       \frac{P(\gamma)}{P(e)} \approx \frac
             {1 + c_\gamma ({\Lambda_s}/{\Lambda_0})^L}
             {1 + c_e ({\Lambda_1}/{\Lambda_0})^L} .
   \label{eq:1D-Psector-Lambda}
   \eeq
where
   \beq
     c_g\equiv 
\frac{[v_s]_g [v_s]_e} {[v_0]_g[v_0]_e}
   \label{eq:1D-cgamma}
   \eeq
where $[v_0]_g [v_0]_e =1/\nG$, for this one-dimensional model.
It follows from \eqr{eq:1D-Psector-Lambda} that 
   \beq
           \ln \frac{P(\gamma)}{P(e)} \approx (c_\gamma- c_e)  e^{-L/\xi_1}
   \label{eq:1D-Psector-xi}
   \eeq
where $\exp(-1/\xi_1) \equiv |\Lambda_s/\Lambda_0| $.
Often $\Lambda_1 <0$; in this case, we must add a factor
$(-1)^{L}$ on the right-hand-side of \eqr{eq:1D-Psector-xi}.
Furthermore, at short $L$, we may see subdominant terms with
shorter decay lengths $\xi_2$ etc., deriving from other eigenvectors
of $T$.

\THINK{It is NOT just representations of the automorphisms, it is
representations of the group itself.  What gives?}

\subsubsection{Example}

\SAVE{A simple example is the cyclic group $Z_n$ (addition modulo $n$)
with $\Sel= \{\pm 1\}$.  Representation $k$  of group element
$g$ is just $\exp (2\pi gk/n)$: you can see that a Fourier transform
really is just one of these representations.  The eigenvalues
are $\Lambda_k = 2 \cos(2 \pi k/n)$.  Thus $\xi_1= n^2/2\pi^2$
in this case, and $c_\gamma \propto - (1-\cos 2\pi \gamma/n)$.}

A useful example is any group $\Group$ when 
$\Sel= \Group \setminus e$, i.e. every element but the
identity is allowed.  In this case 
$T= (1,1,...,1)\otimes (1,1,...,1) - I$.
Thus $\Lambda_0= \nG-1$ and $\Lambda_1=\Lambda_2 =...=-1$.
Thus $\exp(-1/\xi_1)=1/(\nG-1)$ and the deviations have
alternating signs, i.e. the $(-1)^L$ factor is needed 
in \eqr{eq:1D-Psector-xi}.
For the model $Z_2\times Z_2(2)$, the matrix is
  \beq
      T = \begin{pmatrix}
                     1 & 2 & 2 & 2 \\
                     2 & 1 & 2 & 2 \\
                     2 & 2 & 1 & 2 \\
                     2 & 2 & 2 & 1 \end{pmatrix}.
  \eeq

\subsubsection{Symmetry-class reduced matrix}

We can classify eigenvectors as ``symmetric'' or ``asymmetric''
according to what representation of the automorphism group they transform under. 
``Symmetric'' eigenvectors are invariant under group symmetries,
while ``asymmetric'' eigenvectors represent a bias of the 
probability distribution favoring certain
local patterns over other (symmetry-related) ones.

Since the sector probability ratio \eqr{eq:1D-Psector-Lambda} is the same
for all symmetry-related $\gamma$, I believe not only $v-0$ but
also $v_1$ must be totally symmetric.
That affords a considerable simplification, for we can replace $T$ by its
projection $\Ttilde$ onto the group element symmetry classes.
(Such a class consists of elements that map to each under under some
automorphism, so these are at least as large as the conjugacy classes.)
Whereas the dimension of $T$ was the number of group elements $\nG$,
the dimension of $\Ttilde$ is the number of group classes:
$\Ttilde_{ji}$ tells the number of times that $\sigma \gamma$ belongs to
class $j$, if $\gamma$ belongs to class $i$ and $\sigma$ runs over all
$\nS$ elements in $\Sel$.

For example, in the case of the group $A_5$, the matrix is reduced
from $60\times 60$ to $4\times 4$, with entries for elements of
order one (identity), two, three, and five.  (There are two conjugacy
classes with order five, but they are equivalent by an outer 
automorphism.)  For the model $A_5(3)$, we get
  \beq
      \Ttilde = \begin{pmatrix}
                     0 & 0 & 1 & 0 \\
                     0 & 4 & 6 & 5 \\
                    20 & 8 & 7 & 5 \\
                     0 & 8 & 6 & 10 \end{pmatrix}.
  \eeq
This matrix is similar to a symmetric matrix $D^{-1/2}\Ttilde D^{1/2}$;
here $D={\rm diag}(1,15,20,24)$ for this group, or in general is
the diagonal matrix with entries being the count of each class.

\subsection{Sector probabilities in two dimensions?}
\label{sec:2D-TM-sector}

A two-dimensional finite-group height model is also described by 
a transfer matrix $T$.  However, now the vector that $T$ acts on 
represents all possible path products $\gamma_{x,y}$ taken to a point $(x,y)$, 
and thus is $(\nG)^W$ dimensional, where $W$ is the width of the
strip (in the $y$ direction; iteration still runs in the $x$ direction).
We must replace $c_\gamma \to c_\gamma(W)$ and $\xi_1 \to \xi(W)$ in
Eq.~\eqr{eq:1D-Psector-xi}. Conceivably $\xi(W) \to 0$ as $W\to\infty$,
as is very well known in gapless systems, so the form of
$\exp(-L/\xi(W))$ does not prove exponential decay in $d=2$.  

Nevertheless, we can make a plausible guess to obtain a fitting form 
for comparison with numerics.  Since all correlations are expected
to be rapidly decaying, a strip of width $W$  is like $W/w_0$
independent, one-dimensional strips of width $w_0$ in parallel. But all these strips
are constrained to have the same, or equivalent, sector label $\gamma$.)
The consequence is that 
   \beq
       \frac{P(\gamma)}{P(e)} \approx 
     \Bigg[\frac
{1 + c_\gamma ({\Lambda_1}/{\Lambda_0})^L}
{1 + c_e ({\Lambda_1}/{\Lambda_0})^L}
 \Bigg]^{W/w_0}
   \eeq
so 
   \beq
           \ln \frac{P(\gamma)}{P(e)} \approx 
          (C_\gamma-C_e)  W e^{-L/\xi}
   \eeq
in place of \eqr{eq:1D-Psector-xi}, with $C_\gamma \approx c_\gamma/w_0$
and $\xi \approx \xi_1$ independent of $W$.

My chief motivation for introducing the transfer-matrix formalism is
separate from such guesses about the $W$ scaling, and is much better founded.
Namely, the eigenvectors for the corrections to $P(\gamma)$ 
are representations of the automorphism group. Furthermore,
{\it which} representation goes with the longest correlations 
is probably the same as in the one-dimensional case.  
What really matters here is that our choice of a selected set $\Sel$
defines a sort of metric on $\Group$: the distance from $g$ to $g'$
is the number of times you need to 
multiply by an element of $\Sel$ to get from $g$ to $g'$.
Then, the first nontrivial eigenvector $v_1$ is the 
mode that is slowest varying on $\Group$ according to this 
metric (apart from $v_0$ which is uniform).

The one-dimensional correlation length $\xi_1$ can be computed for any
combination of $\Group$ and $\Sel$ and can serve as another
``figure of merit'' for a group.  That is, in light of the
previous paragraph's argument, it should be roughly
related to the true sector probability decay length $\xi$
for the two-dimensional model (and likely related to the 
defect-defect decay length as well).

\subsection{Defect separations and $W=2$ transfer matrix}
\label{sec:W3-TM-defects}

Whereas the one-dimensional model already seems to capture
the essence of how sector probabilities depend on system
size and sector label, it does not admit topological defects 
and hence sheds no light on the parallel question of 
how $p(R)$ for a defect pair decays with separation or
depends on the respective defect charges.

Clearly, $p(R)$ must be associated somehow with the eigenvectors
and eigenvalues of the two-dimensional transfer matrix,
since all possible correlation information is expressed in it.
But it is not self-evident just what kind of distortion of
the ensemble is being propagated, or what sort of subdominant
eigenvector: the symmetric kind
(which governed the sector probabilities) or the 
asymmetric kind.

I will work out here a toy calculation, again using a transfer matrix,
of the correlation decay due to {\it asymmetric} eigenvectors.
I believe they are the ones that matter for the case of an {\it abelian}
group.  In that case, the charge of a defect is a particular element:
the loop product around the defect gives that same result, no matter
how big the loop, and only another defect with the inverse of that
charge can cancel it.  In  the non-abelian case, however, these
properties would seem to be defined only modulo conjugacy classes.
Therefore, the picture presented here is only asserted to go with
abelian groups.

The simplest property that could influence or be influenced by
a defect's presence is the correlation of two adjacent spins on the
same plaquette, i.e.  sitting on bonds that make a 90$^\circ$ angle.
\SAVE{This expectation is reinforced by the analogy to the well-understood
case of vortex defects in an XY model -- defects in height models are
analogous. In that case, the group elements correspond to angle gradients 
and the curl of this is related to presence of vortices.}
A simple example is the model $Z_2\times Z_2(2)$sq, in which
$\nS=3$ elements are allowed -- all except the identity.
These elements are $\{a,b,c\}$.
Consider a plaquette with the spins on two edges specified
and the remaining two spins to be assigned 
(there are $3^2$ unconstrained ways to do so).
When adjacent edges on a plaquette have the same element, there
are three ways to satisfy the plaquette constraint, but only
two ways if the given adjacent edges are different.  
On the other hand, if we want to make a defect plaquette,
there are six ways when the given spins are the same but
seven ways when they are different.

Let's set up a $W=2$ strip, the narrowest
kind that can capture defect correlations.
This transfer matrix, unlike the previous one,
refers to the actual spin configurations in each
vertical pair of bonds; we add up all the possible
horizontal bonds.  
I assume the upper row of plaquettes are constrained 
to be have identity product around the plaquette.
Plaquettes in the lower row can have any product -- 
defects are permitted -- with a weight $\theta_0$ for
the identity or $\theta_a,\theta_b,\theta_c$ for
the respective defect charges $a, b, c$.  We imagine the limit
in which $\theta_{a,b,c}$ are small and ask for the
corresponding defect correlations.

Although $T$ has $3^2 \times 3^2 = 81$ elements,
in fact there are only ten distinct kinds by symmetry,
as given in Table~\ref{tab:Z2-W2-TM}; 
``no.'' represents the number of times each kind
occurs in the matrix.
The factors $\theta_0\approx 1$ and $\theta_\sigma \ll 1$
are omitted in the table.
To compute the matrix elements, note that
when $\sigma_1=\sigma_1'$ in the upper plaquette,
the central horizontal bond may be any element
[three possibilities] but if $\sigma_1\neq \sigma_1'$,
the central bond may be only $\sigma_1$ or $\sigma_1'$
[two possibilities].
There are always three possibilities for the lower horizontal
bond, so the table's rows add up to 9 or 6
depending whether or not $\sigma_1=\sigma_1'$.

The probability to find a defect of charge $\beta'$ at separation $R$, 
given there is a defect of charge $\beta$ at the origin, is then
\beq
    p(R) = \frac {\Tr\Big( [T^{(0)}]^{L-R-1}
   T^{(\beta')} [T^{(0)}]^{R-1} T^{(\beta)}  \Big)}
     { \Tr\Big( [T^{(0)}]^{L-1} T^{(\beta)} \Big) }
\eeq
For a large power $M$, we can replace 
    $[T^{(0)}]^M  \to (\Lambda_0)^M v_0 \otimes v_0  +
                  (\Lambda_1)^M v_1 \otimes v_1 $.
Here $v_0$ and $v_1$ are the eigenvectors belonging to the maximum and
next-largest eigenvalues of $T^{(0)}$.
Assuming $L\gg R \gg 1$, we get
  \begin{align}
    p(R) &=  \frac{ \Lambda_0^{L-R-1} 
                \sum _{k=0,1} \sum_m (v_0,T^{\beta'} v_{k,m}) (v_{k,m}T^\beta v_0) \Lambda_{k,m}^{R-1} }
              { \Lambda_0^{L-1} (v_0,T^{\beta} v_0)}  \\
         &= p_0(\beta') \Big[1  + c(\beta') c(\beta) 
                           \Big(\frac{\Lambda_1}{\Lambda_0}\Big)^R \Big] 
  \label{eq:pR-Lambda}
  \end{align}
where 
 \beq
       p_0(\beta')= \frac{(v_0, T^{(\beta')}v_0) }{\Lambda_0}
 \eeq
--- remember $(v_0,T^{(0)} v_0)= \Lambda_0$] ---
and
  \beq
       c(\beta) \equiv \Big(\frac{\Lambda_0}{\Lambda_1}\Big)^{1/2}
            \frac{ (v_1,T^{(\beta)} v_0)} {(v_0,T^{(\beta)} v_0)}.
  \eeq

\begin{table}
\caption{Example for group $Z_2\times Z_2$(2):
Transfer matrix elements $T^{(\beta)}_{\sigma_1',\sigma_0';\sigma_1,\sigma_0}$}
\begin{tabular}{|lll| rrrr |}
\hline 
no. & $(\sigma_1,\sigma_0)$   &  $(\sigma_1',\sigma_0')$  & 
     $T^{(0)}$ & $T^{(a)}$ & $T^{(b)}$ & $T^{(c)}$ \\
\hline 
3  & $(aa)$ & $(aa)$ &  3  &  2 & 2 & 2 \\
12 & $(aa)$ & $(ab)$ &  2  &  2 & 2 & 3 \\
12 & $(aa)$ & $(ba)$ &  2  & 1 & 1 & 2 \\
 6 & $(aa)$ & $(bb)$ &  2  & 1 & 1  & 2 \\
12 & $(aa)$ & $(bc)$ &  1  &  2 & 2& 1 \\
6  & $(ab)$ & $(ab)$ &  3  &  2 & 2 & 2 \\
6  & $(ab)$ & $(ba)$ &  2  & 1 & 1 & 2 \\
6 & $(ab)$ & $(ac)$ &   2 & 3 & 2 & 2 \\
6 & $(ba)$ & $(ca)$ &  2 & 2 & 1 & 1 \\
12 & $(ab)$ & $(ca)$ & 1 & 2 & 1 & 2 \\
\hline 
\end{tabular}
\label{tab:Z2-W2-TM}
\end{table}
Please remember, the eigenvector called $v_1$ here is asymmetric, and
is thus not the same as the symmetric eigenvector called $v_s$ in
Sec.~\ref{sec:1D-TM-sector}.  We see that asymptotically,
  \beq
     \ln p(R)  \propto c(\beta)c(\beta') e^{-R/\xi}
    \label{eq:defect-p}
  \eeq
Notice first that the decay length $\xi$ is independent of the defect
charges, but different defect charges have different projections onto
this eigenmode.  As is clear from the derivation, a more general form
could be written, including subdominant contributions:
  \beq
     \ln p(R)  \propto \sum _k c_k(\beta)c_k(\beta') e^{-R/\xi_k}
  \eeq
where $\xi_1>\xi_2 >...$.  The later terms could be important 
corrections to include in fits at short $R$, particularly when 
the smaller $\xi_k$'s happen to be associated 
with larger coefficients $c_k (\beta)$.  Also, if $c_1(\beta)=0$ for
certain defects, their asymptotic interaction gets carried by the
first mode that has nonzero projections onto both defects.

The formulas basically apply to any width of strip.
(If defects are allowed in more than one
horizontal row of plaquettes, then the defect distribution
is no longer a function just of $R$ but also of the two
$y$ coordinates;  the only modification necessary is 
that $T^{(\beta)} \to T^{(\beta;y)}$, labeled not only
by the defect's flavor but by its $y$ coordinate.)
I would speculate that the $W=3$ strip, with defects in the central
row, may be a good approximation in practice, although of course
there is no control parameter to make small.  The basis for this
is simply the notion that, when we have rapid exponential decays,
these are associated in the ensemble with strings connecting 
the defects; any influence carried by a less direct chain would
be exponentially smaller in correspondence with the longer
length.

\subsection{Other approaches to $p(R)$ in $d=2$}

I conjecture there is an alternative approach which is more
congenial to $d=2$. 
Namely, in the vicinity of a defect, the probabilities
of local patterns have small deviations from the bulk
values, which could be represented by operators $O_k(\rr)$ and
small conjugate fields $h_k(\rr)$.
That is, adding a Hamiltonian $\sum _\rr h_k(\rr) O_k(\rr)$
(in the absence of the defect) would perturb the ensemble 
the same way as the defect does.  Note that 
the operator ``$O_k(\rr)$'' is schematic, in the sense that
such operators probably involve two spins at different $\rr$
(in light of the same logic laid out in the first paragraphs of this subsection).
Then possibly some sort of mean-field approximation produces a difference equation
for $h_k(\rr)$, the discrete analog of Poisson's equation
$\nabla^2 h_k(\rr) =  h_k(\rr)/\xi_k^2$, which has solutions
$\sim e^{-R/\xi_k}/R$.   
In this approach,  we have a sort of small parameter in
that the influence of a perturbation decays as $e^{-R/\xi_k}$,
which  becomes arbitrary small at sufficiently large $R$.
We can therefore rely on linear response in that regime.

One can conceive additional approaches to $p(R)$ which depend
on a genuine small parameter;  the difficulty is that the
actual model families defined in this paper are far from that
limit.  For example, one could expand around the pure gauge
theory: in place of the spin constraint, there would be no
constraint but configurations would have a statistical
weight $\exp(\lambda \sum _{\rr,\rr'} u(\sigma(\rr,\rr'))$, where
$u(\sigma)$ would penalize all $\sigma\notin\Sel$.  
In the limit $\lambda\to 0$, we have a pure gauge theory
in which all correlation lengths $\xi_k$ are zero, so
hopefully $\xi_k$ would scale as a power of $\lambda$.
The models under consideration are, unfortunately, the
case $\lambda=\infty$. Still, since the topological phases
are like the pure gauge models at large scales, they
should be adiabatically connected and hence this approach 
should be qualitatively valid.

\section{Discussion}
\label{sec:discussion}

I have put forward the notion of purely classical 
topological order, defined by an ergodicity breaking into 
sectors dependent on the topology, and not distinguishable
by thermodynamic expectations of any local operator.
A family of explicit models has been described, along with a
suitable Monte Carlo technique, and criteria were suggested 
to pick out the most promising cases (in having a nontrivial
and updatable ensemble of allowed states).

A framework was set up (Sec.~\ref{sec:defns})
to define models with three variable attributes: 
which group, which class(es) out
of the group to selected as ``spin'' variables, and which
lattice to place the model on.  These are characterized
by parameters -- the sizes $\nG$ and $\nS$ of the group and 
the spin subset, the coordination number $z$ of the lattice 
and $z_d$ of its dual -- which entered crude formulas 
that estimate the entropy of the model and its updatability
under single-site Monte Carlo moves (Sec.~\ref{sec:criteria-models}
and ~\ref{sec:criteria-update}), which are the only actual calculations 
in the paper.  Groups with normal subgroups tend to be
``less nonabelian '', thus perhaps less attractive 
(Sec.~\ref{sec:example-models}).  Although topological order
superficially would appear to be intrinsically featureless, 
there is sufficient richness of measurable functions when 
one considers the dependence of free energy on
topological indices -- finite size dependence on sector
or finite distance dependence on defect separations
(Sec.~\ref{sec:measurements}).

In trying to connect the classical picture to the 
quantum theory of topological order, it is intriguing
that a given two defect  charges (see Sec.~\ref{sec:defects}) 
can combine in more than one way, in a classical non-abelian
model, reminiscent of fusion rules in a quantum model.
If further investigation finds that the sector counting 
gives the same degeneracies in the classical as in the
quantum case, one would conclude that this is one of
the shared properties, not an intrinsically quantum one.

The physical manifestations of classical topological order 
and/or of non-abelianness
are less striking, perhaps, than for the quantum case.
Most prominent is the behavior of topological defects.
Topological order implies deconfinement in the classical model
for nonabelian and abelian cases alike.
Non-abelianness (of the  group) changes the rules for addition of defect
charges, and braiding has physical consequences, even though
there are no Berry phases in a classical model. (It must be
noted, however, that the same behaviors are seen in 
non-abelian defects of ordinary long-range order~\cite{mermin}
-- they are not inherent to topological order.)

Degeneracies of different topological sectors, the defining property
of topological order, work differently in the
non-abelian than in the abelian case: for example, there are far distinct 
fewer sectors in the non-abelian case (Sec.~\ref{sec:sectors}).

\subsection{Quantum mechanics}

Several central concepts of topological order are inherently 
quantum-mechanical and thus have {\it no} counterpart in 
classical topological order.   They are mainly related to
phases in wavefunctions and braiding of worldlines in
2+1 dimensions, namely anyon and mutual statistics.
Most real or imagined experiments relating to topological 
excitations have involved interference phenomena 
(e.g. tunneling in various geometries of quantum Hall fluids)  
and thus probe the quantum-mechanical aspects of topological
order. 

Another feature missing in the classical models is the
dual defect or quasiparticle (such as the vison~\cite{senthil}), 
which is a distortion of the phase factors in the many-body
wavefunction.

A final attribute of topological orders is the nontrivial
counting statistics of the excited states made by several
quasiparticles, which is quantum mechanical in that it
concerns the linear dimension of a Hilbert space.
One cannot rule out the appearance of similar concepts
in classical stat mech -- there, too, the partition function
contains combinatorial factors for the placement of 
defects, after the other degrees of freedome have been 
integrated out.  However, I am not aware of a classical
situation in which such a nontrivial counting actually emerges.

\subsection{Construction of quantum models?}

Any of the classical height models with topological order
may be converted into a simular {\it quantum} model  
if we can endow it ``flipping'' move, just as classical
dimer (and other) models get converted into quantum dimer models
using the Rokhsar-Kivelson (RK) prescription~\cite{RK}.
A barrier to this is that the only generally guaranteed
``flip'' move is a cluster update, as explained in
Sec.~\ref{sec:cluster-update}.
\SAVE{Conceivably a finite ``atlas'' of different candidate
cluster updates would suffice to access all the states.  
But even that would be too messy to write as a Hamiltonian.}

Fortunately, whenever the single-site update (Sec. \ref{sec:singlesite}) 
suffices, we {\it can} define a quantum model with a simple ``flipping'' term in
the Hamiltonian, usually parametrized by an amplitude $t$, as well as
a ``potential'' term of strength $V=t$ that penalizes each flippable
place.   At the RK point $V=t$, the ground state wavefunction is a 
superposition of all configurations in the same topological sector,
with the same (equal) weighting as in the classical ensemble, and
(mutually inaccessible) topological sectors are trivially degenerate.  
One is also free to set $V=0$ --  obtaining a simpler model in which
flippable sites are so favored that an ordered state is likely
to be the outcome -- or to vary $V/t$ with the hope of crossing
a phase transition. 

The above recipe is incomplete, in that there are many possible choices 
of update (labeled by the multiplier $\tau$ of Sec.~\ref{sec:MC}), and
presumably all or many should be included in 
``flipping'' term of the quantum Hamiltonian, which requires a 
prescription for the relative magnitudes of coefficient to put for each
class of $\tau$, as well as the relative phase factors.
Presumably, a proper choice is taking the same phase factors for
every term, i.e. the Hamiltonian transforms by the fully symmetric 
(trivial) representation of the automorphism group of $\Group$.
\SAVE{I incorrectly thought that requires giving up the full group symmetry.
An partial example of that can be seen in the
transverse field (quantum) Ising model. Choosing a
particular factor for the down-to-up term
(and its conjugate for the up-to-down term)
amounts to picking the field's orientation
in the transverse plane. In this special case, it does 
not actually hurt the classical $Z_2$ symmetry, it 
only hurts the additional rotation symmetry that quantum
spins should have.}
Alternatively, in lucky cases, one might select a site-dependent 
pattern of $\tau_i$'s so as to link the group symmetry to the lattice symmetry,
in the spirit of Kitaev's honeycomb model~\cite{Ki03}.  Another 
option is to include a second, quantum fluctuating field of
$\tau's$ which are used for the update. 
If the $\tau's$ are derived 
from a second set of ``spins'' also having the gauge-like structure of 
a finite-group height model, we might be able to realize dual
(``magnetic'') defects having mutual statistics with the $\sigma$-spin type 
(``electric'') defects 
described in this paper.

\subsection{Possible simulations}

As laid out in Sec.~\ref{sec:measurements},
several quantities e.g. correlation functions can be measured in classical 
non-abelian height models as a test-bed, whereas the analogous 
calculation might be very challenging computationally in a
quantum mechanical model.   Of course, the answers need not be the same,
but the questions may be much clearer once the classical
results are in hand.
First, one can create defect pairs and
evaluate the histogram of their separations, which 
will reveal whether or not they are deconfined.
Secondly, one can evaluate the probabilities of
different topological sectors, which is the direct test 
of topological order.

Furthermore, if we generalize the models to include
classical Hamiltonians (so as to weight configurations
according to the Boltzmann distribution), 
phase transitions can be studied.
Just as a standard height model may have a
``smooth'' phase, in which one or more height
components becomes locked, it seems conceivable
that a discrete-group height model might have a phase in
which the loop products can take values in a subgroup
$\Group' \subset \Group$ .
If so, one might encounter critical points separating
different topological phases, and characterize the
critical exponents.

\subsection{Dilution and effective interactions of
local degrees of freedom?}

A classical model might be a helpful too for 
investigating the consequences of dilution disorder
in a model with topological order.
Each diluted site or plaquette is like a very small hole
cut in the system, thereby increasing the genus and
the number of topological sectors.  If the hole were big, these
sectors would be truly degenerate (by the definition of 
topological order), however this degeneracy is broken
since the holes are small.   

The values of $\gamma^*$ on each dilution site are local
pseudospins, which are expected to 
have (exponentially decaying) interactions mediated 
by the fluid in between them, like the emergent
spin-1/2 degrees of freedom in diluted spin-1 
antiferromagnetic chains~\cite{hagiwara90}.
The ground state of such a 
system could be constructed by a renormalization group
that iteratively combines the most strongly coupled 
pair of pseudospins into a single effective pseudospin, 
as was originally done for the (exponentially decaying) antiferromagnetic coupling 
of the charge-bound electron spins in P-doped  Si~\cite{bhatt82}.

In the non-abelian case, at least, we do not know for sure
whether these interactions lead to an inert singlet phase
(as in the antiferromagnet) or could give a state with 
some kind of order among the pseudospins. 
Thus the system with defects would {\it not} be a topological liquid,
and this would be a novel scenario of how order can emerge
due to disorder.~\cite{wessell00}

\acknowledgments
I thank M. Troyer for suggesting the problem;
also L. Ioffe,  A. Kitaev, D. A. Ivanov,  Simon Trebst,
C. Castelnovo, C. Chamon, S. Papanikolaou, and R. Lamberty
for comments and discussion.
ALso, I thank J. Papaioannou and R. Maimon
for preliminary work on the simulation algorithm.
This work was supported by NSF Grant No. DMR-1005466.

\end{document}